# Fast Transceiver Design for RIS-Assisted MIMO mmWave Wireless Communications

Haiyue Jing, *Student Member, IEEE*, Wenchi Cheng, *Senior Member, IEEE*, Xiang-Gen Xia, *Fellow, IEEE*

***Abstract*—Due to high bandwidth and small antenna size, millimeter-wave (mmWave) integrated line-of-sight (LOS) multiple-input-multiple-output (MIMO) systems have attracted much attention. Reconfigurable intelligent surfaces (RISs), which have the potential to change the characteristics of incident electromagnetic waves with low power cost, can improve the performance for the MIMO mmWave wireless communications. Uniform circular array (UCA) is an effective antenna structure with low complexity transceiver. In this paper, UCA based RIS-assisted MIMO mmWave wireless communications with transmit UCA, the RIS UCAs, and receive UCA are investigated. Since the rotation angles between the transceiver make the channel matrix noncirculant, an algorithm is developed to derive the ranges of the rotation angles based on an acceptable error and reduce the impact of rotation angles on channel matrix. Then, we propose a low-complexity precoding scheme at the transmitter, phase designs at the RIS UCAs, and a phase compensation scheme at the receiver, which can convert the channel matrix into an equivalent circulant channel matrix with a small error. Then, a fast symbol-wise maximum likelihood (ML) detection scheme is proposed to recover the signals with low computational complexity. Simulation results are presented to illustrate the theory.**

## I. Introduction

Millimeter-wave (mmWave) communication, which is considered as a powerful technique for wireless communications operating in the 30 to 300 GHz range, has attracted much attention during the past years [1]–[5]. It has been shown that mmWave multiple-input-multiple-output (MIMO) systems can significantly increase the capacity of wireless communications due to the abundance of available frequency bandwidth [1]–[3] and there have been many related studies in the literature. However, there are some challenges for mmWave wireless communications [6]–[9]. One of the challenges is that there exist severe attenuations for signals in mmWave wireless communications. Also, mmWave wireless communication systems suffer from high penetration losses and the high directivity at mmWave bands makes mmWave wireless communication systems much more sensitive to signal blockage, resulting in the wireless channel over mmWave band usually has no paths [10].

Recently, reconfigurable intelligent surfaces (RISs), as an emerging technology for the future sixth generation (6G) wireless communication network, have been proposed to overcome some of the shortcomings of non-line-of-sight (non-LOS) mmWave channels [11]–[15]. RIS is a planar meta-surface with multiple low-cost passive reflecting elements and each element can be controlled by adjusting the amplitudes and/or phases of the incident signals [12]–[20]. An important application of RISs is to provide transmission paths to achieve high-capacity transmission when the direct transmission is blocked [21]–[24].

However, for the RIS-assisted MIMO mmWave wireless communications, most works focus on capacity enhancement and there are few papers taking the computational complexity reduction into consideration. They mostly use uniform linear arrays (ULAs) at transmitters and receivers, which may have high demodulation complexity using maximum likelihood (ML) detection [25]–[27]. In contrast to ULAs, uniform circular arrays (UCAs), say $N$ antenna elements, have been studied for MIMO systems [28]–[30], where orthogonal frequency division multiplexing (OFDM) signals across antenna direction are transmitted so that at the receiver side, the fast symbol-wise maximum likelihood (ML) demodulation can be achieved similar to the conventional uncoded OFDM systems, and in the meantime, $N$ symbols per channel use are achieved. The key idea is that the channel matrix between transmit and receive UCAs can be formulated as a circulant matrix similar to the conventional OFDM system over an intersymbol interference (ISI) channel.

For the two UCAs at transmitter and receiver, there may exist relative rotation angles along $x$, $y$, and $z$-axes. If these rotation angles are large, there is an impact on the channel phase, resulting in that it is impossible for the channel matrix to be converted into a circulant matrix. However, when these rotation angles are small, their impact on the channel phase can be ignored. The key idea of this paper is to place multiple RISs all with UCAs between transmitter and receiver so that for each segment, the relative rotation angles are small. Thus, the overall rotation angles between transmitter and receiver are divided into smaller relative rotation angles for each segment. Some simple phase compensations on RIS UCAs are proposed and can help the channel matrix of each segment to be approximated as a circulant matrix with a small error. Then, the overall channel matrix from transmitter to receiver, which is the product of channel matrices from all the segments,

This work was supported in part by the National Key Research and Development Program of China under Grant 2021YFC3002102 and in part by the Key R&D Plan of Shaanxi Province under Grant 2022ZDLGY05-09.
*(Corresponding author: Wenchi Cheng.)*

H. Jing and W. Cheng are with the State Key Laboratory of Integrated Services Networks, Xidian University, Xi'an 710071, China (e-mail: hyjing@stu.xidian.edu.cn; wccheng@xidian.edu.cn).

X.-G. Xia is with Xidian University, Xi'an 710071, China, and also with the Department of Electrical and Computer Engineering, University of Delaware, Newark, DE 19716 USA (e-mail: xxia@ee.udel.edu).

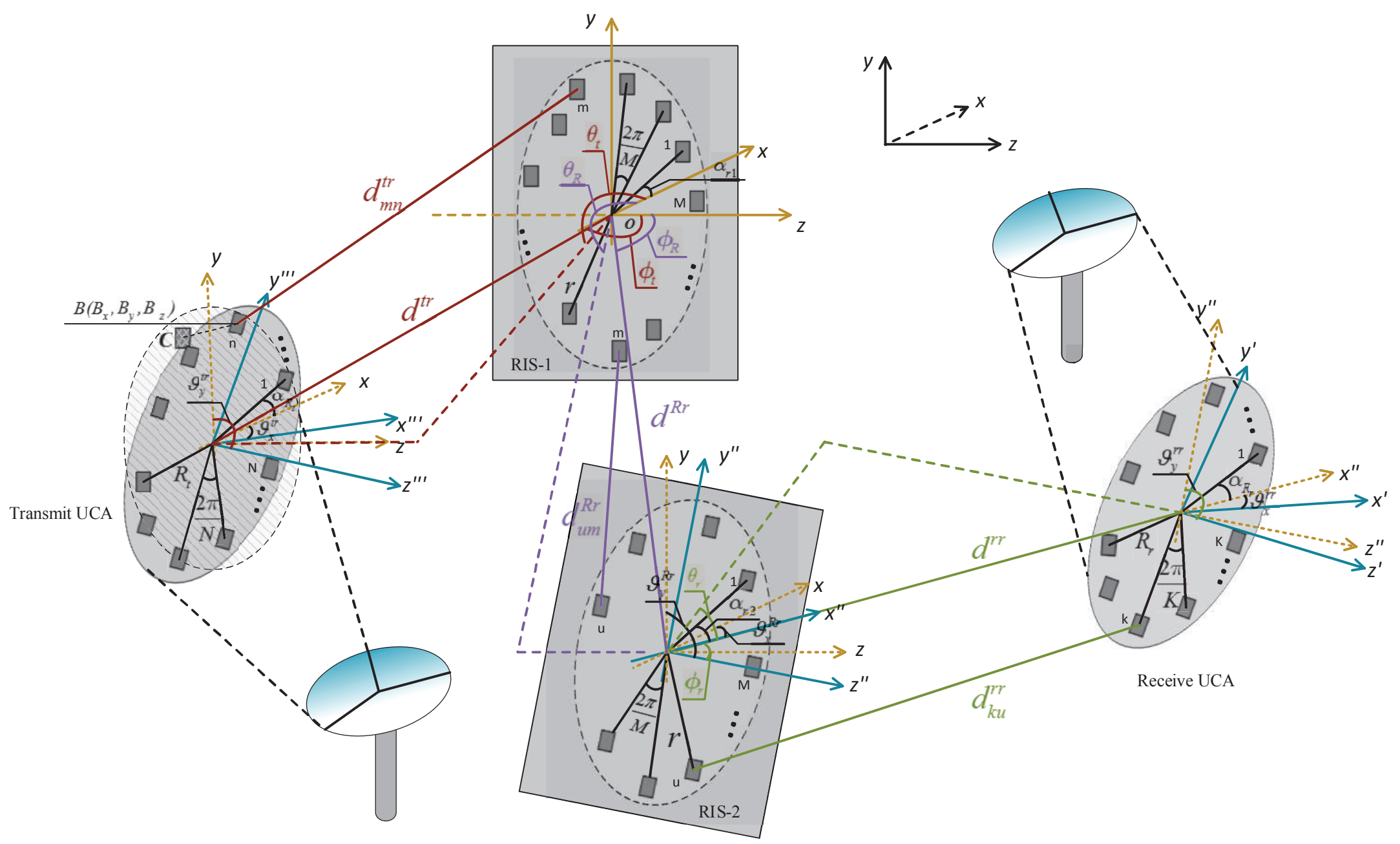

Fig. 1. System model for the UCA based RIS-assisted MIMO system with misaligned transmit UCA, RIS UCAs, and receive UCA.

can be considered as a circulant matrix. The error of the approximation for the overall channel matrix with RIS UCAs is much smaller than that without RIS UCAs. Although, more RIS UCAs can be applied between transmitter and receiver, for convenience in this paper we only consider two RIS UCAs. We assume that the channel on each segment is line-of-sight (LOS) and also assume that there are no signals across different segment channels, for example, the signals from the transmitter can only be received by the nearest RIS and cannot be received by any other RISs or the receiver.

Using our proposed low complexity precoding scheme on the transmitter, phase designs on the RIS UCAs, and phase compensation scheme on the receiver, the overall channel matrix can converted to an equivalent circulant matrix with an acceptable error. Then, a fast symbol-wise ML detection is proposed on the receiver, which has a low computational complexity. Simulations results are then presented to illustrate our proposed fast transceiver design for RIS assisted MIMO systems.

The rest of this paper is organized as follows. Section II gives the system model for the UCA based RIS-assisted MIMO system where the transmit UCA, the RIS UCAs, and the receive UCA may not be aligned with each other. Section III derives the channel models from the transmit UCA to RIS-1 UCA, from RIS-1 UCA to RIS-2 UCA, from RIS-2 UCA to the receive UCA, and from the transmitter to the receiver. Also, an algorithm is developed to derive the ranges of rotation angles for each segment based on an acceptable error level. Section IV proposes a low complexity precoding scheme on the transmitter, phase designs on the RIS UCAs, and a phase compensation scheme and a fast symbol-wise ML detection scheme on the receiver. Section V analyzes the bit error rate (BER) and the computational complexity. Section VI shows some simulation results to illustrate our proposed transceiver. The paper is concluded in Section VII.

## II. System Model

Fig. 1 depicts the system model for a UCA based and RIS-assisted MIMO system with misaligned transmit UCA, two RIS UCAs, and receive UCA. The signals emitted by the transmit UCA are first received and reflected by RIS-1 UCA. Then, RIS-2 UCA reflects the signals from RIS-1 UCA and the receive UCA finally receives the signals reflected by RIS-2 UCA. We assume that there are no paths from the transmit UCA to RIS-2 UCA, or to the receiver, and no paths from RIS-1 UCA to the receive UCA. Since the signals via multiple reflections between the RISs can be ignored at the receiver due to small signal-to-noise-ratios (SNRs) [31], [32], the multiple reflections between RIS-1 UCA and RIS-2 UCA are not considered in this paper. Two RIS UCAs are placed with small rotation angles based on the positions of the transmitter and the receiver. Therefore, the large rotation angles between the transmitter and the receiver can be divided into some small rotation angles using the RIS UCAs in the middle of the transmit UCA and receive UCA[1]. Note that, for convenience, this paper only considers two RIS UCAs, while more RIS UCAs can be similarly discussed if needed.

Denote by $R_t$, $r_1$, $r_2$, and $R_r$ the radius of the transmit UCA, the radius of RIS-1 UCA, the radius of RIS-2 UCA, and the radius of the receive UCA, respectively. Denote by $d^{tr}$, $d^{Rr}$, and $d^{rr}$ the distance from the center of the transmit UCA to the center of RIS-1 UCA, the distance between the centers of RIS-1 UCA and RIS-2 UCA, and the distance between the center of RIS-2 UCA and the center of the receive UCA, respectively. The transmit UCA, the two RIS UCAs, and the receive UCA are equipped with $N$ antenna elements, $M$ array elements, and $K$ antenna elements, respectively. In this paper we use $N = K = M$ to ensure that the overall channel matrix can be converted into a circulant matrix. Denote by $\alpha_{R_t}$, $\alpha_{r1}$, and $\alpha_{r2}$ the angles between the first antenna elements and the direction with zero radian, i.e., the $x$-axis, corresponding to the transmit UCA, RIS-1 UCA, and RIS-2 UCA, respectively. Denote by $\alpha_{R_r}$ the angle between the first antenna element and the direction with zero radian, i.e., the $x''$-axis, corresponding to the receive UCA. For the transmit UCA, RIS-1 UCA, and RIS-2 UCA, the coordinate system, denoted by $xyz$, is on the RIS-1 UCA plane. While for the RIS-2 UCA and the receive UCA, the coordinate system, denoted by $x''y''z''$, is on the RIS-2 UCA plane. The distances between any two points (antenna elements or array elements) at the adjacent UCAs for each segment in different coordinate systems are the same.

[1]The impact of the included angles can be neglected using our proposed transceiver design. Thus, we do not discuss the included angles in this paper.

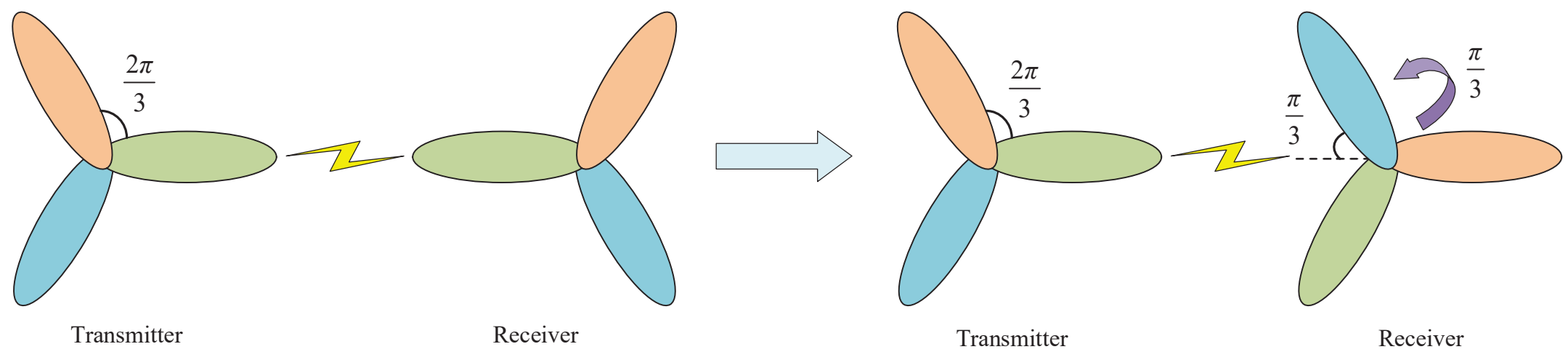

Fig. 2. Radiation profiles of the transmitter and receiver.

For the coordinate system $xyz$, the parameter $\theta_t$ denotes the included angle between $x$-axis and the projection of the line from the center of RIS-1 UCA to the center of the transmit UCA on the RIS-1 UCA plane. The notation $\theta_R$ is the included angle between $x$-axis and the projection of the line from the center of RIS-1 UCA to the center of RIS-2 UCA on the RIS-1 UCA plane. Also, $\phi_t$ denotes the included angle between $z$-axis and the line from the center of RIS-1 UCA to the center of the transmit UCA. We denote by $\phi_R$ the included angle between $z$-axis and the line from the center of RIS-1 UCA to the center of RIS-2 UCA. The rotation angles $\vartheta_y^{tr}$ and $\vartheta_y^{Rr}$ represent the angles of rotation around the $x$-axis of the transmit UCA and RIS-2 UCA, respectively. The rotation angles $\vartheta_x^{tr}$ and $\vartheta_x^{Rr}$ represent the angles of rotation around the $y$-axis of the transmit UCA and RIS-2 UCA, respectively.

For the coordinate system $x''y''z''$, we denote by $\theta_r$ the included angle between $x''$-axis and the projection of the line from the center of RIS-2 UCA to the center of the receive UCA on the RIS-2 UCA plane. The parameter $\phi_r$ denotes the included angle between $z''$-axis and the line from the center of RIS-2 UCA to the center of the receive UCA. We denote by $\vartheta_y^{rr}$ and $\vartheta_x^{rr}$ the angles of rotation around the $x''$-axis and $y''$-axis corresponding to the receive UCA, respectively. Note that the above two coordinate systems are used to calculate all the distances and angles in each segment of the system, which determine the channel in each segment.

Fig. 2 shows the radiation profiles of the transmitter and the receiver. There are three sectors at the transmitter and receiver, respectively, and the angles between any two sectors are $2\pi/3$. When the transmitter and the receiver are misaligned, there exist rotation angles between the transmitter and the receiver. Since the rotation angles between any two sectors at the transmitter and the receiver, respectively, are $2\pi/3$, the rotation angles between the transmitter and the receiver are smaller than $\pi/3$. Then, when the angles of two sectors are smaller than $\pi/3$, the two sectors are selected at the transmitter and the receiver to achieve high-efficient transmission. Thus, we can consider that the rotation angles between the transmitter and the receiver is not larger than $\pi/3$. For convenience, we consider that the rotation angles, $\vartheta_y^{tr}$, $\vartheta_x^{tr}$, $\vartheta_y^{Rr}$, $\vartheta_x^{Rr}$, $\vartheta_y^{rr}$, and $\vartheta_x^{rr}$ in the misaligned system are not larger than $\pi/9$ and Section III-D gives the details.

## III. Channel Model

### A. Channel Model From the Transmitter to RIS-1 UCA

Let $h_{mn}^{tr}$ and $d_{mn}^{tr}$ denote the channel gain and the distance, respectively, from the $n$th, $1 \leq n \leq N$, antenna element on the transmitter to the $m$th, $1 \leq m \leq N$, array element on the RIS-1 UCA. Then, $h_{mn}^{tr}$ can be written as follows:

$$h_{mn}^{tr} = \frac{\beta \lambda e^{-j\frac{2\pi}{\lambda} d_{mn}^{tr}}}{4\pi d_{mn}^{tr}}, \tag{1}$$

where $\beta$ denotes the combination of all the relevant constants such as attenuation and phase rotation caused by antennas and their patterns on both sides.

From the transmit UCA to the RIS-1 UCA, the coordinate of the $m$th array element on the RIS-1 UCA is

$$(r_1 \cos(\varphi_m + \alpha_{r1}), r_1 \sin(\varphi_m + \alpha_{r1}), 0),$$

where $\varphi_m = 2\pi(m-1)/M$ is the rotation-angle for the RIS-1 UCA and $\varphi_m + \alpha_r$ is the azimuthal angle, defined as the angular position on a plane perpendicular to the axis of propagation, of the $m$th array element on the RIS-1 UCA. The coordinate of the center of the transmit UCA is

$$(d^{tr} \sin\phi_t \cos\theta_t, d^{tr} \sin\phi_t \sin\theta_t, d^{tr} \cos\phi_t),$$

where $d^{tr}$ represents the distance between the center of the transmit UCA and the center of RIS-1 UCA.

To derive the coordinate of the $n$th antenna element on the transmit UCA for the coordinate system $xyz$, two operations, which are rotation and translation, are used. For the rotation operation, the coordinate of the $n$th antenna element on the transmit UCA, denoted by $(B_x', B_y', B_z')$, is as follows:

$$\begin{cases} \mathcal{E}_n^{tr} = d^{tr} R_t \sin\phi_t \cos\theta_t \cos(\psi_n + \alpha_{R_t}) \cos\vartheta_y^{tr} + d^{tr} R_t \sin\phi_t \sin\theta_t \sin(\psi_n + \alpha_{R_t}) \cos\vartheta_x^{tr} \\ \qquad + d^{tr} R_t \sin\phi_t \cos\theta_t \sin(\psi_n + \alpha_{R_t}) \sin\vartheta_x^{tr} \sin\vartheta_y^{tr} + d^{tr} R_t \cos\phi_t \cos(\psi_n + \alpha_{R_t}) \sin\vartheta_y^{tr} \\ \qquad - d^{tr} R_t \cos\phi_t \sin(\psi_n + \alpha_{R_t}) \sin\vartheta_x^{tr} \cos\vartheta_y^{tr}; \quad (8a) \\ \mathcal{D}_{mn}^{tr} = -R_t r_1 \cos(\psi_n + \alpha_{R_t}) \cos(\varphi_m + \alpha_{r1}) \cos\vartheta_y^{tr} - R_t r_1 \sin(\psi_n + \alpha_{R_t}) \sin(\varphi_m + \alpha_{r1}) \cos\vartheta_x^{tr} \\ \qquad - R_t r_1 \cos(\psi_n + \alpha_{R_t}) \sin(\varphi_m + \alpha_{r1}) \sin\vartheta_x^{tr} \sin\vartheta_y^{tr} \\ = \underbrace{-\frac{R_t r_1}{2}(1 + \cos\vartheta_y^{tr} \cos\vartheta_x^{tr}) \cos(\psi_n - \varphi_m + \alpha_{R_t} - \alpha_{r1} - \chi_1^{tr})}_{\hat{\mathcal{D}}_{mn}^{tr}} \underbrace{- \frac{R_t r_1}{2}(1 - \cos\vartheta_y^{tr} \cos\vartheta_x^{tr}) \cos(\psi_n + \varphi_m + \alpha_{R_t} + \alpha_{r1} - \chi_2^{tr})}_{\widetilde{\mathcal{D}}_{mn}^{tr}}; \quad (8b) \\ \mathcal{F}_m^{tr} = -d^{tr} r_1 \sin\phi_t \cos(\varphi_m + \alpha_{r1} - \theta_t). \quad (8c) \end{cases}$$

$$\begin{bmatrix} B'_x \\ B'_y \\ B'_z \end{bmatrix} = [\boldsymbol{R}_x(\vartheta_y^{tr}) \boldsymbol{R}_y(\vartheta_x^{tr})]^T \begin{bmatrix} R_t \cos(\psi_n + \alpha_{R_t}) \\ R_t \sin(\psi_n + \alpha_{R_t}) \\ 0 \end{bmatrix} \tag{2}$$

where $\psi_n = 2\pi(n-1)/N$ is the rotation-angle and $\psi_n + \alpha_{R_t}$ is the azimuthal angle for the $n$th antenna element on the transmit UCA, $(\cdot)^T$ represents the transpose operation, and $\boldsymbol{R}_x(\vartheta_y^{tr})$ and $\boldsymbol{R}_y(\vartheta_x^{tr})$ represent the direction cosine matrices corresponding to the $x$-axis and $y$-axis, respectively:

$$\boldsymbol{R}_x(\vartheta_y^{tr}) = \begin{bmatrix} 1 & 0 & 0 \\ 0 & \cos\vartheta_y^{tr} & \sin\vartheta_y^{tr} \\ 0 & -\sin\vartheta_y^{tr} & \cos\vartheta_y^{tr} \end{bmatrix} \tag{3}$$

and

$$\boldsymbol{R}_y(\vartheta_x^{tr}) = \begin{bmatrix} \cos\vartheta_x^{tr} & 0 & -\sin\vartheta_x^{tr} \\ 0 & 1 & 0 \\ \sin\vartheta_x^{tr} & 0 & \cos\vartheta_x^{tr} \end{bmatrix}, \tag{4}$$

respectively.

With the translation of the center

$$(d^{tr} \sin\phi_t \cos\theta_t, d^{tr} \sin\phi_t \sin\theta_t, d^{tr} \cos\phi_t),$$

the coordinate of the $n$th antenna element, denoted by $(B_x, B_y, B_z)$, on the transmit UCA for the coordinate system $xyz$ is as follows:

$$\begin{bmatrix} B_x \\ B_y \\ B_z \end{bmatrix} = \begin{bmatrix} B'_x \\ B'_y \\ B'_z \end{bmatrix} + \begin{bmatrix} d^{tr} \sin\phi_t \cos\theta_t \\ d^{tr} \sin\phi_t \sin\theta_t \\ d^{tr} \cos\phi_t \end{bmatrix}. \tag{5}$$

Then, $B_x$, $B_y$, and $B_z$ are

$$\begin{cases} B_x = d^{tr} \sin\phi_t \cos\theta_t + R_t \cos(\psi_n + \alpha_{R_t}) \cos\vartheta_y^{tr} \\ \qquad + R_t \sin(\psi_n + \alpha_{R_t}) \sin\vartheta_x^{tr} \sin\vartheta_y^{tr}; \\ B_y = d^{tr} \sin\phi_t \sin\theta_t + R_t \sin(\psi_n + \alpha_{R_t}) \cos\vartheta_x^{tr}; \\ B_z = d^{tr} \cos\phi_t - R_t \cos(\psi_n + \alpha_{R_t}) \sin\vartheta_y^{tr} \\ \qquad + R_t \sin(\psi_n + \alpha_{R_t}) \sin\vartheta_x^{tr} \cos\vartheta_y^{tr}. \end{cases} \tag{6}$$

The distance $d_{mn}^{tr}$ between the $n$th antenna element on the transmit UCA and the $m$th array element on the RIS-1 UCA can be derived as:

$$\begin{aligned} d_{mn}^{tr} &= \\ &\sqrt{[B_x - r_1 \cos(\varphi_m + \alpha_{r1})]^2 + [B_y - r_1 \sin(\varphi_m + \alpha_{r1})]^2 + B_z^2} \\ &= \sqrt{(d^{tr})^2 + R_t^2 + r_1^2 + 2\mathcal{E}_n^{tr} + 2\mathcal{D}_{mn}^{tr} + 2\mathcal{F}_m^{tr}}, \end{aligned} \tag{7}$$

where $\mathcal{E}_n^{tr}$, $\mathcal{D}_{mn}^{tr}$ and $\mathcal{F}_m^{tr}$ are given in (8).

In (8), $\chi_1^{tr}$ and $\chi_2^{tr}$ are defined as follows:

$$\begin{cases} \cos\chi_1^{tr} = \frac{\cos\vartheta_y^{tr} + \cos\vartheta_x^{tr}}{1 + \cos\vartheta_y^{tr} \cos\vartheta_x^{tr}}; \\ \sin\chi_1^{tr} = \frac{-\sin\vartheta_y^{tr} \sin\vartheta_x^{tr}}{1 + \cos\vartheta_y^{tr} \cos\vartheta_x^{tr}}, \end{cases} \tag{9}$$

and

$$\begin{cases} \cos\chi_2^{tr} = \frac{\cos\vartheta_y^{tr} - \cos\vartheta_x^{tr}}{1 - \cos\vartheta_y^{tr} \cos\vartheta_x^{tr}}; \\ \sin\chi_2^{tr} = \frac{\sin\vartheta_y^{tr} \sin\vartheta_x^{tr}}{1 - \cos\vartheta_y^{tr} \cos\vartheta_x^{tr}}. \end{cases} \tag{10}$$

Because of $d^{tr} \gg r_1$ and $d^{tr} \gg R_t$, we can obtain that $\frac{\mathcal{E}_n^{tr} + \mathcal{D}_{mn}^{tr} + \mathcal{F}_m^{tr}}{\sqrt{(d^{tr})^2 + R_t^2 + r_1^2}}$ is close to 0 and we can make approximation for $d_{mn}^{tr}$ in the denominator and numerator of (1). The distance $d_{mn}^{tr}$ can be rewritten according to $\sqrt{1+2x} \approx 1 + x$ when $x$ is close to zero as follows:

$$\begin{aligned} d_{mn}^{tr} &= \sqrt{(d^{tr})^2 + R_t^2 + r_1^2} \sqrt{1 + \frac{2\mathcal{E}_n^{tr} + 2\mathcal{D}_{mn}^{tr} + 2\mathcal{F}_m^{tr}}{(d^{tr})^2 + R_t^2 + r_1^2}} \\ &\approx \sqrt{(d^{tr})^2 + R_t^2 + r_1^2} + \frac{\mathcal{E}_n^{tr} + \mathcal{D}_{mn}^{tr} + \mathcal{F}_m^{tr}}{\sqrt{(d^{tr})^2 + R_t^2 + r_1^2}}. \end{aligned} \tag{11}$$

To obtain $h_{mn}^{tr}$, we need the following lemma.

**Lemma 1.**

$$\frac{1}{d_{mn}^{tr}} \approx \frac{1}{\sqrt{(d^{tr})^2 + R_t^2 + r_1^2}}, \tag{12}$$

*Proof.* See Appendix A. □

$$\begin{cases} \mathcal{E}_u^{Rr} = d^{Rr} r_2 \sin\phi_R \cos\theta_R \cos(\psi_u + \alpha_{r2}) \cos\vartheta_y^{Rr} + d^{Rr} r_2 \sin\phi_R \sin\theta_R \sin(\psi_u + \alpha_{r2}) \cos\vartheta_x^{Rr} \\ \qquad + d^{Rr} r_2 \sin\phi_R \cos\theta_R \sin(\psi_u + \alpha_{r2}) \sin\vartheta_x^{Rr} \sin\vartheta_y^{Rr} - d^{Rr} r_2 \cos\phi_R \cos(\psi_u + \alpha_{r2}) \sin\vartheta_y^{Rr} \\ \qquad + d^{Rr} r_2 \cos\phi_R \sin(\psi_u + \alpha_{r2}) \sin\vartheta_x^{Rr} \cos\vartheta_y^{Rr}; \quad (19a) \\ \mathcal{D}_{um}^{Rr} = -r_1 r_2 \cos(\psi_u + \alpha_{r2}) \cos(\varphi_m + \alpha_{r1}) \cos\vartheta_y^{Rr} - r_1 r_2 \sin(\psi_u + \alpha_{r2}) \sin(\varphi_m + \alpha_{r1}) \cos\vartheta_x^{Rr} \\ \qquad - r_1 r_2 \sin(\psi_u + \alpha_{r2}) \cos(\varphi_m + \alpha_{r1}) \sin\vartheta_x^{Rr} \sin\vartheta_y^{Rr} \\ = \underbrace{-\frac{r_1 r_2}{2}(1 + \cos\vartheta_y^{Rr} \cos\vartheta_x^{Rr}) \cos(\psi_u - \varphi_m + \alpha_{r2} - \alpha_{r1} - \chi_1^{Rr})}_{\hat{\mathcal{D}}_{um}^{Rr}} \underbrace{- \frac{r_1 r_2}{2}(1 - \cos\vartheta_y^{Rr} \cos\vartheta_x^{Rr}) \cos(\psi_u + \varphi_m + \alpha_{r2} + \alpha_{r1} - \chi_2^{Rr})}_{\widetilde{\mathcal{D}}_{um}^{Rr}}; \\ \qquad (19b) \\ \mathcal{F}_m^{Rr} = -d^{Rr} r_1 \sin\phi_R \cos(\varphi_m + \alpha_{r1} - \theta_R). \quad (19c) \end{cases}$$

Then, using the above-mentioned approximation, we can rewrite $h_{mn}^{tr}$ in (1):

$$h_{mn}^{tr} \approx \frac{\beta\lambda e^{-j\frac{2\pi}{\lambda}\sqrt{(d^{tr})^2 + R_t^2 + r_1^2}}}{4\pi\sqrt{(d^{tr})^2 + R_t^2 + r_1^2}} e^{-j\frac{2\pi}{\lambda}\frac{\mathcal{E}_n^{tr} + \mathcal{D}_{mn}^{tr} + \mathcal{F}_m^{tr}}{\sqrt{(d^{tr})^2 + R_t^2 + r_1^2}}}. \quad (13)$$

The channel matrix, denoted by $\boldsymbol{H}_{tr}$ with element $h_{mn}^{tr}$, from the transmit UCA to RIS-1 UCA can be written as:

$$\boldsymbol{H}_{tr} = \boldsymbol{U}_2 \widetilde{\boldsymbol{H}}_{tr} \boldsymbol{U}_1, \quad (14)$$

where $\boldsymbol{U}_1$ and $\boldsymbol{U}_2$ are

$$\begin{cases} \boldsymbol{U}_1 = \mathrm{diag}\left(e^{-j\frac{2\pi}{\lambda}\frac{\mathcal{E}_1^{tr}}{\sqrt{(d^{tr})^2 + R_t^2 + r_1^2}}}, \cdots, e^{-j\frac{2\pi}{\lambda}\frac{\mathcal{E}_N^{tr}}{\sqrt{(d^{tr})^2 + R_t^2 + r_1^2}}}\right); \\ \boldsymbol{U}_2 = \mathrm{diag}\left(e^{-j\frac{2\pi}{\lambda}\frac{\mathcal{F}_1^{tr}}{\sqrt{(d^{tr})^2 + R_t^2 + r_1^2}}}, \cdots, e^{-j\frac{2\pi}{\lambda}\frac{\mathcal{F}_M^{tr}}{\sqrt{(d^{tr})^2 + R_t^2 + r_1^2}}}\right). \end{cases} \quad (15)$$

The notation $\widetilde{\boldsymbol{H}}_{tr}$ in (14) is a matrix with its $m$th row and $n$th column element $\widetilde{h}_{mn}^{tr}$, which is derived as:

$$\begin{aligned} \widetilde{h}_{mn}^{tr} &= \frac{\beta\lambda e^{-j\frac{2\pi}{\lambda}\sqrt{(d^{tr})^2 + R_t^2 + r_1^2}}}{4\pi\sqrt{(d^{tr})^2 + R_t^2 + r_1^2}} e^{-j\frac{2\pi}{\lambda}\frac{\mathcal{D}_{mn}^{tr}}{\sqrt{(d^{tr})^2 + R_t^2 + r_1^2}}} \\ &= \frac{\beta\lambda e^{-j\frac{2\pi}{\lambda}\sqrt{(d^{tr})^2 + R_t^2 + r_1^2}}}{4\pi\sqrt{(d^{tr})^2 + R_t^2 + r_1^2}} e^{-j\frac{2\pi}{\lambda}\frac{\hat{\mathcal{D}}_{mn}^{tr} + \widetilde{\mathcal{D}}_{mn}^{tr}}{\sqrt{(d^{tr})^2 + R_t^2 + r_1^2}}}, \end{aligned} \quad (16)$$

where $\hat{\mathcal{D}}_{mn}^{tr}$ and $\widetilde{\mathcal{D}}_{mn}^{tr}$ are given in (8b).

### B. Channel Model From RIS-1 UCA to RIS-2 UCA

In general, the power consumption on the RIS is pratically negligible. Then, the power of the incident signals keeps unchanged on the RIS. After modifying the phases of the incident signals, RIS reflects the incident signals [14]. Let $h_{um}^{Rr}$ and $d_{um}^{Rr}$ denote the channel gain and the distance, respectively, from the $m$th array element on the RIS-1 UCA to the $u$th, $1 \le u \le N$, array element on the RIS-2 UCA. Then, $h_{um}^{Rr}$ can be written as:

$$h_{um}^{Rr} = \frac{\beta\lambda e^{-j\frac{2\pi}{\lambda}d_{um}^{Rr}}}{4\pi d_{um}^{Rr}}. \quad (17)$$

Similar to the channel gain $h_{mn}^{tr}$, $h_{um}^{Rr}$ can be derived as:

$$h_{um}^{Rr} \approx \frac{\beta\lambda e^{-j\frac{2\pi}{\lambda}\sqrt{(d^{Rr})^2 + r_1^2 + r_2^2}}}{4\pi\sqrt{(d^{Rr})^2 + r_1^2 + r_2^2}} e^{-j\frac{2\pi}{\lambda}\frac{\mathcal{E}_u^{Rr} + \mathcal{D}_{um}^{Rr} + \mathcal{F}_m^{Rr}}{\sqrt{(d^{Rr})^2 + r_1^2 + r_2^2}}}, \quad (18)$$

where $\mathcal{E}_u^{Rr}$, $\mathcal{D}_{um}^{Rr}$, and $\mathcal{F}_m^{Rr}$ are given in (19).

The notations $\chi_1^{Rr}$ and $\chi_2^{Rr}$ in (19) are defined as follows:

$$\begin{cases} \cos\chi_1^{Rr} = \frac{\cos\vartheta_y^{Rr} + \cos\vartheta_x^{Rr}}{1 + \cos\vartheta_y^{Rr}\cos\vartheta_x^{Rr}}; \\ \sin\chi_1^{Rr} = \frac{\sin\vartheta_y^{Rr}\sin\vartheta_x^{Rr}}{1 + \cos\vartheta_y^{Rr}\cos\vartheta_x^{Rr}}, \end{cases} \quad (20)$$

and

$$\begin{cases} \cos\chi_2^{Rr} = \frac{\cos\vartheta_y^{Rr} - \cos\vartheta_x^{Rr}}{1 - \cos\vartheta_y^{Rr}\cos\vartheta_x^{Rr}}; \\ \sin\chi_2^{Rr} = \frac{\sin\vartheta_y^{Rr}\sin\vartheta_x^{Rr}}{1 - \cos\vartheta_y^{Rr}\cos\vartheta_x^{Rr}}. \end{cases} \quad (21)$$

The channel matrix, denoted by $\boldsymbol{H}_{Rr}$, from the RIS-1 UCA to the RIS-2 UCA is given as:

$$\boldsymbol{H}_{Rr} = \boldsymbol{U}_4 \widetilde{\boldsymbol{H}}_{Rr} \boldsymbol{U}_3, \quad (22)$$

where $\boldsymbol{U}_3$ and $\boldsymbol{U}_4$ are

$$\begin{cases} \boldsymbol{U}_3 = \mathrm{diag}\left(e^{-j\frac{2\pi}{\lambda}\frac{\mathcal{F}_1^{Rr}}{\sqrt{(d^{Rr})^2 + r_1^2 + r_2^2}}}, \cdots, e^{-j\frac{2\pi}{\lambda}\frac{\mathcal{F}_M^{Rr}}{\sqrt{(d^{Rr})^2 + r_1^2 + r_2^2}}}\right); \\ \boldsymbol{U}_4 = \mathrm{diag}\left(e^{-j\frac{2\pi}{\lambda}\frac{\mathcal{E}_1^{Rr}}{\sqrt{(d^{Rr})^2 + r_1^2 + r_2^2}}}, \cdots, e^{-j\frac{2\pi}{\lambda}\frac{\mathcal{E}_M^{Rr}}{\sqrt{(d^{Rr})^2 + r_1^2 + r_2^2}}}\right), \end{cases} \quad (23)$$

and are also unitary. The $u$th row and $m$th column element $\widetilde{h}_{um}^{Rr}$ of the matrix $\widetilde{\boldsymbol{H}}_{Rr}$ in (22) is

$$\begin{aligned} \widetilde{h}_{um}^{Rr} &= \frac{\beta\lambda e^{-j\frac{2\pi}{\lambda}\sqrt{(d^{Rr})^2 + r_1^2 + r_2^2}}}{4\pi\sqrt{(d^{Rr})^2 + r_1^2 + r_2^2}} e^{-j\frac{2\pi}{\lambda}\frac{\mathcal{D}_{um}^{Rr}}{\sqrt{(d^{Rr})^2 + r_1^2 + r_2^2}}} \\ &= \frac{\beta\lambda e^{-j\frac{2\pi}{\lambda}\sqrt{(d^{Rr})^2 + r_1^2 + r_2^2}}}{4\pi\sqrt{(d^{Rr})^2 + r_1^2 + r_2^2}} e^{-j\frac{2\pi}{\lambda}\frac{\hat{\mathcal{D}}_{um}^{Rr} + \widetilde{\mathcal{D}}_{um}^{Rr}}{\sqrt{(d^{Rr})^2 + r_1^2 + r_2^2}}}, \end{aligned} \quad (24)$$

where $\hat{\mathcal{D}}_{um}^{Rr}$ and $\widetilde{\mathcal{D}}_{um}^{Rr}$ are given in (19b).

$$\begin{cases} \mathcal{E}_k^{rr} = d^{rr} R_r \sin\phi_r \cos\theta_r \cos(\psi_k + \alpha_{R_r}) \cos\vartheta_y^{rr} + d^{rr} R_r \sin\phi_r \sin\theta_r \sin(\psi_k + \alpha_{R_r}) \cos\vartheta_x^{rr} \\ \qquad + d^{rr} R_r \sin\phi_r \cos\theta_r \sin(\psi_k + \alpha_{R_r}) \sin\vartheta_x^{rr} \sin\vartheta_y^{rr} - d^{rr} R_r \cos\phi_r \cos(\psi_k + \alpha_{R_r}) \sin\vartheta_y^{rr} \\ \qquad + d^{rr} R_r \cos\phi_r \sin(\psi_k + \alpha_{R_r}) \sin\vartheta_x^{rr} \cos\vartheta_y^{rr}; \quad (27a) \\ \mathcal{D}_{ku}^{rr} = -R_r r_2 \cos(\psi_k + \alpha_{R_r}) \cos(\varphi_u + \alpha_{r2}) \cos\vartheta_y^{rr} - R_r r_2 \sin(\psi_k + \alpha_{R_r}) \sin(\varphi_u + \alpha_{r2}) \cos\vartheta_x^{rr} \\ \qquad - R_r r_2 \sin(\psi_k + \alpha_{R_r}) \cos(\varphi_u + \alpha_{r2}) \sin\vartheta_x^{rr} \sin\vartheta_y^{rr} \\ = \underbrace{-\frac{R_r r_2}{2}(1+\cos\vartheta_y^{rr}\cos\vartheta_x^{rr})\cos(\psi_k - \varphi_u + \alpha_{R_r} - \alpha_{r2} - \chi_1^{rr})}_{\hat{\mathcal{D}}_{ku}^{rr}} \underbrace{-\frac{R_r r_2}{2}(1-\cos\vartheta_y^{rr}\cos\vartheta_x^{rr})\cos(\psi_k + \varphi_u + \alpha_{R_r} + \alpha_{r2} - \chi_2^{rr})}_{\widetilde{\mathcal{D}}_{ku}^{rr}}; \quad (27b) \\ \mathcal{F}_u^{rr} = -d^{rr} r_2 \sin\phi_r \cos(\varphi_u + \alpha_{r2} - \theta_r). \quad (27c) \end{cases}$$

### C. Channel Model From RIS-2 UCA to the Receiver

Let $h_{ku}^{rr}$ and $d_{ku}^{rr}$ denote the channel gain and the distance, respectively, from the $u$th array element on the RIS-2 UCA to the $k$th, $1 \leq k \leq N$, antenna element on the receiver. Then, $h_{ku}^{rr}$ can be written as:

$$h_{ku}^{rr} = \frac{\beta\lambda e^{-j\frac{2\pi}{\lambda}d_{ku}^{rr}}}{4\pi d_{ku}^{rr}}. \tag{25}$$

Similar to the channel gain $h_{mn}^{tr}$, $h_{ku}^{rr}$ can be derived as:

$$h_{ku}^{rr} \approx \frac{\beta\lambda e^{-j\frac{2\pi}{\lambda}\sqrt{(d^{rr})^2+R_r^2+r_2^2}}}{4\pi\sqrt{(d^{rr})^2+R_r^2+r_2^2}} e^{-j\frac{2\pi}{\lambda}\frac{\mathcal{E}_k^{rr}+\mathcal{D}_{ku}^{rr}+\mathcal{F}_u^{rr}}{\sqrt{(d^{rr})^2+R_r^2+r_2^2}}}, \tag{26}$$

where $\mathcal{E}_k^{rr}$, $\mathcal{D}_{ku}^{rr}$, and $\mathcal{F}_u^{rr}$ are given in (27).

The notations $\chi_1^{rr}$ and $\chi_2^{rr}$ in (27) are defined as follows:

$$\begin{cases} \cos\chi_1^{rr} = \frac{\cos\vartheta_y^{rr}+\cos\vartheta_x^{rr}}{1+\cos\vartheta_y^{rr}\cos\vartheta_x^{rr}}; \\ \sin\chi_1^{rr} = \frac{\sin\vartheta_y^{rr}\sin\vartheta_x^{rr}}{1+\cos\vartheta_y^{rr}\cos\vartheta_x^{rr}}, \end{cases} \tag{28}$$

and

$$\begin{cases} \cos\chi_2^{rr} = \frac{\cos\vartheta_y^{rr}-\cos\vartheta_x^{rr}}{1-\cos\vartheta_y^{rr}\cos\vartheta_x^{rr}}; \\ \sin\chi_2^{rr} = \frac{\sin\vartheta_y^{rr}\sin\vartheta_x^{rr}}{1-\cos\vartheta_y^{rr}\cos\vartheta_x^{rr}}. \end{cases} \tag{29}$$

The channel matrix, denoted by $\boldsymbol{H}_{rr}$, from RIS-2 UCA to the receive UCA is given as:

$$\boldsymbol{H}_{rr} = \boldsymbol{U}_6 \widetilde{\boldsymbol{H}}_{rr} \boldsymbol{U}_5, \tag{30}$$

where $\boldsymbol{U}_5$ and $\boldsymbol{U}_6$ are

$$\begin{cases} \boldsymbol{U}_5 = \mathrm{diag}\left(e^{-j\frac{2\pi}{\lambda}\frac{\mathcal{F}_1^{rr}}{\sqrt{(d^{rr})^2+R_r^2+r_2^2}}}, \cdots, e^{-j\frac{2\pi}{\lambda}\frac{\mathcal{F}_M^{rr}}{\sqrt{(d^{rr})^2+R_r^2+r_2^2}}}\right); \\ \boldsymbol{U}_6 = \mathrm{diag}\left(e^{-j\frac{2\pi}{\lambda}\frac{\mathcal{E}_1^{rr}}{\sqrt{(d^{rr})^2+R_r^2+r_2^2}}}, \cdots, e^{-j\frac{2\pi}{\lambda}\frac{\mathcal{E}_K^{rr}}{\sqrt{(d^{rr})^2+R_r^2+r_2^2}}}\right). \end{cases} \tag{31}$$

The $k$th row and $u$th column element $\widetilde{h}_{ku}^{rr}$ of the matrix $\widetilde{\boldsymbol{H}}_{rr}$ is

$$\begin{aligned} \widetilde{h}_{ku}^{rr} &= \frac{\beta\lambda e^{-j\frac{2\pi}{\lambda}\sqrt{(d^{rr})^2+R_r^2+r_2^2}}}{4\pi\sqrt{(d^{rr})^2+R_r^2+r_2^2}} e^{-j\frac{2\pi}{\lambda}\frac{\mathcal{D}_{ku}^{rr}}{\sqrt{(d^{rr})^2+R_r^2+r_2^2}}} \\ &= \frac{\beta\lambda e^{-j\frac{2\pi}{\lambda}\sqrt{(d^{rr})^2+R_r^2+r_2^2}}}{4\pi\sqrt{(d^{rr})^2+R_r^2+r_2^2}} e^{-j\frac{2\pi}{\lambda}\frac{\hat{\mathcal{D}}_{ku}^{rr}+\widetilde{\mathcal{D}}_{ku}^{rr}}{\sqrt{(d^{rr})^2+R_r^2+r_2^2}}}, \end{aligned} \tag{32}$$

where $\hat{\mathcal{D}}_{ku}^{rr}$ and $\widetilde{\mathcal{D}}_{ku}^{rr}$ are given in (27b).

### D. The Overall Channel Model of the System

The overall channle model (from the transmitter to the receiver), denoted by $\boldsymbol{H}$, of the system is $\boldsymbol{H}_{rr}\boldsymbol{H}_2^r\boldsymbol{H}_{Rr}\boldsymbol{H}_1^r\boldsymbol{H}_{tr}$, where $\boldsymbol{H}_1^r$ and $\boldsymbol{H}_2^r$ are the phase compensation matrices on RIS-1 UCA and RIS-2 UCA, respectively, which will be given in (46) and (49) in Subsection IV-B. Then, we have

$$\begin{aligned} \boldsymbol{H} &= \boldsymbol{H}_{rr}\boldsymbol{H}_2^r\boldsymbol{H}_{Rr}\boldsymbol{H}_1^r\boldsymbol{H}_{tr} \\ &= \boldsymbol{U}_6\widetilde{\boldsymbol{H}}_{rr}\boldsymbol{U}_5\boldsymbol{U}_5^*\boldsymbol{U}_4^*\boldsymbol{U}_4\widetilde{\boldsymbol{H}}_{Rr}\boldsymbol{U}_3\boldsymbol{U}_3^*\boldsymbol{U}_2^*\boldsymbol{U}_2\widetilde{\boldsymbol{H}}_{tr}\boldsymbol{U}_1 \\ &= \boldsymbol{U}_6\widetilde{\boldsymbol{H}}_{rr}\widetilde{\boldsymbol{H}}_{Rr}\widetilde{\boldsymbol{H}}_{tr}\boldsymbol{U}_1. \end{aligned} \tag{33}$$

The items $\widetilde{\mathcal{D}}_{mn}^{tr}$, $\widetilde{\mathcal{D}}_{um}^{Rr}$, and $\widetilde{\mathcal{D}}_{ku}^{rr}$ in (8b), (19b), and (27b), respectively, render the channel matrix noncirculant. This is because the directions of cyclic shifts for $\widetilde{\mathcal{D}}_{mn}^{tr}$ related to $m$ and $n$, $\widetilde{\mathcal{D}}_{um}^{Rr}$ related to $u$ and $m$, and $\widetilde{\mathcal{D}}_{ku}^{rr}$ related to $k$ and $u$ are left, while the directions of cyclic shifts for $\hat{\mathcal{D}}_{mn}^{tr}$, $\hat{\mathcal{D}}_{um}^{Rr}$, and $\hat{\mathcal{D}}_{ku}^{rr}$ are right. When there is a small rotation, i.e., the rotation angles are small and the items $\widetilde{\mathcal{D}}_{mn}^{tr}$, $\widetilde{\mathcal{D}}_{um}^{Rr}$, and $\widetilde{\mathcal{D}}_{ku}^{rr}$ can be neglected. Then, the overall channel matrix can be proved to be circulant using Lemma 2. We denote by $\widehat{\boldsymbol{H}}_{tr}$, $\widehat{\boldsymbol{H}}_{Rr}$, and $\widehat{\boldsymbol{H}}_{rr}$ the channel matrices from the transmit UCA to RIS-1 UCA, from RIS-1 UCA to RIS-2 UCA, and from RIS-2 UCA to the receive UCA, respectively, after the procoding on the transmit UCA, phase compensations on the two RISs, and the phase compensation on the receive UCA when $\widetilde{\mathcal{D}}_{mn}^{tr}$, $\widetilde{\mathcal{D}}_{um}^{Rr}$, and $\widetilde{\mathcal{D}}_{ku}^{rr}$ are regarded as 0. Denote by $\widehat{h}_{mn}^{tr}$, $\widehat{h}_{um}^{Rr}$, and $\widehat{h}_{ku}^{rr}$ the elements of $\widehat{\boldsymbol{H}}_{tr}$, $\widehat{\boldsymbol{H}}_{Rr}$, and $\widehat{\boldsymbol{H}}_{rr}$, respectively. Let $\epsilon$ denote the error between the channel model and approximated channel model from the transmit UCA to the RIS-1 UCA as follows:

$$\epsilon \triangleq \frac{\sum_{m=1}^{M}\sum_{n=1}^{N}|\widetilde{h}_{mn}^{tr} - \widehat{h}_{mn}^{tr}|^2}{\sum_{m=1}^{M}\sum_{n=1}^{N}|\widetilde{h}_{mn}^{tr}|^2}, \tag{34}$$

where $\widehat{h}_{mn}^{tr}$ is

$$\widehat{h}_{mn}^{tr} = \frac{\beta\lambda e^{-j\frac{2\pi}{\lambda}\sqrt{(d^{tr})^2+R_t^2+r_1^2}}}{4\pi\sqrt{(d^{tr})^2+R_t^2+r_1^2}} e^{-j\frac{2\pi}{\lambda}\frac{\hat{\mathcal{D}}_{mn}^{tr}}{\sqrt{(d^{tr})^2+R_t^2+r_1^2}}}. \tag{35}$$

Fig. 3 illustrates the error $\epsilon$ versus the rotation angles $\vartheta_x^{tr}$ and $\vartheta_y^{tr}$. As shown in Fig. 3, the graph is symmetrical with respect to $\vartheta_x^{tr} = \vartheta_y^{tr}$. Also, we can observe that the error $\epsilon$ is close to zero when the rotation angles $\vartheta_x^{tr}$ and $\vartheta_y^{tr}$ are small. Therefore, we can consider the channel matrix is circulant when the rotation angles $\vartheta_x^{tr}$ and $\vartheta_y^{tr}$ are small.

$$\widehat{h}_{kn} = \frac{\beta^3 \lambda^3 e^{-j\frac{2\pi}{\lambda}\left(\sqrt{(d^{rr})^2+R_r^2+r_2^2}+\sqrt{(d^{Rr})^2+r_1^2+r_2^2}\sqrt{(d^{tr})^2+R_t^2+r_1^2}\right)}}{64\pi^3\sqrt{(d^{rr})^2+R_r^2+r_2^2}\sqrt{(d^{Rr})^2+r_1^2+r_2^2}\sqrt{(d^{tr})^2+R_t^2+r_1^2}} \sum_{m=1}^{N}\sum_{u=1}^{N} e^{-j\frac{2\pi}{\lambda}\left(\frac{\hat{\mathcal{D}}_{ku}^{rr}}{\sqrt{(d^{rr})^2+R_r^2+r_2^2}}+\frac{\hat{\mathcal{D}}_{um}^{Rr}}{\sqrt{(d^{Rr})^2+r_1^2+r_2^2}}+\frac{\hat{\mathcal{D}}_{mn}^{tr}}{\sqrt{(d^{tr})^2+R_t^2+r_1^2}}\right)}. \quad (37)$$

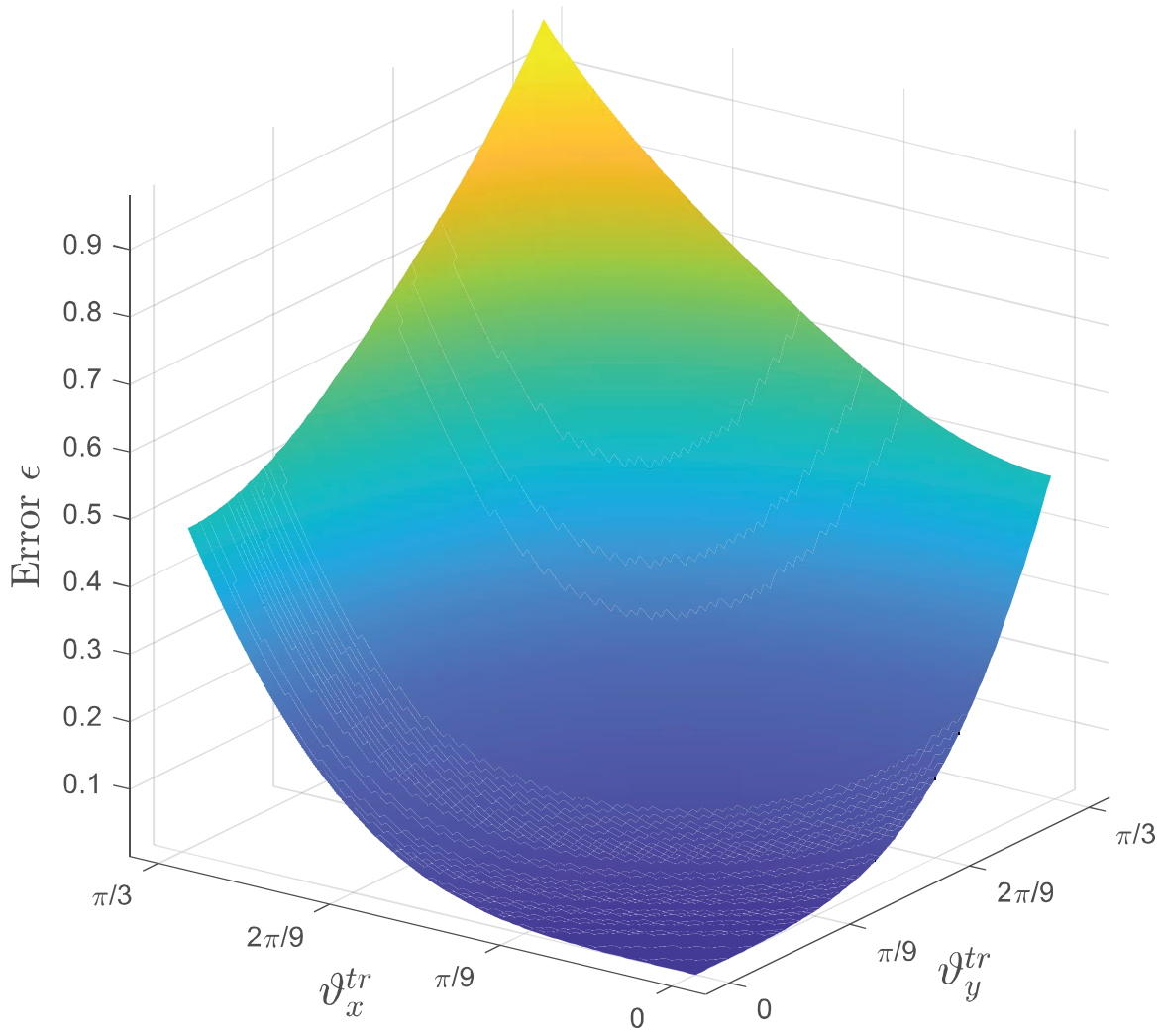

Fig. 3. The error $\epsilon$ versus the rotation angles $\vartheta_x^{tr}$ and $\vartheta_y^{tr}$.

We set a threshold $\varrho$, which represents an acceptable error threshold, and the ranges of rotation angles can be obtained according to $\varrho$. When $\epsilon < \varrho$, i.e., the rotation angles are within the obtained ranges based on $\varrho$, the items $\widetilde{\mathcal{D}}_{mn}^{tr}$, $\widetilde{\mathcal{D}}_{um}^{Rr}$, and $\widetilde{\mathcal{D}}_{ku}^{rr}$ in (8b), (19b), and (27b), respectively, can be neglected. An algorithm, which is developed to find the ranges of rotation angles for a given threshold $\varrho$, is presented in Algorithm 1. For example, after the precoding on the transmit UCA, phase compensations on the two RISs, and the phase compensation on the receive UCA in Section IV, we can obtain that the rotation angles ranges from 0 to $\pi/9$ when $\varrho = 10^{-4}$ with fixed distance, radii, and included angles using our proposed algorithm and the error increases as the rotation angles increase.

When the rotation angles for each segment are in the rotation angle ranges obtained according to $\varrho$, we can consider the rotation angles for each segment are small enough and the impact of the rotation angles on the channel gain can be ignored. Since the rotation angles $\vartheta_y^{tr}$, $\vartheta_x^{tr}$, $\vartheta_y^{Rr}$, $\vartheta_x^{Rr}$, $\vartheta_y^{rr}$, and $\vartheta_x^{rr}$ are small for all the segments, the items $\widetilde{\mathcal{D}}_{mn}^{tr}$, $\widetilde{\mathcal{D}}_{um}^{Rr}$, and $\widetilde{\mathcal{D}}_{ku}^{rr}$ in (8b), (19b), and (27b), respectively, can be considered as 0. Then, for the UCA based RIS-assisted MIMO system, we denote by $\epsilon^t$ the error between the overall channel model and the overall approximated channel model from the transmit UCA to the receive UCA as follows:

$$\epsilon^t = \frac{\| \widetilde{\boldsymbol{H}}_{rr}\widetilde{\boldsymbol{H}}_{Rr}\widetilde{\boldsymbol{H}}_{tr} - \widehat{\boldsymbol{H}}_{rr}\widehat{\boldsymbol{H}}_{Rr}\widehat{\boldsymbol{H}}_{tr} \|^2}{\| \widetilde{\boldsymbol{H}}_{rr}\widetilde{\boldsymbol{H}}_{Rr}\widetilde{\boldsymbol{H}}_{tr} \|^2}. \quad (36)$$

Fig. 4 shows the error between the channel model and the approximated channel model from the transmit UCA to the receive UCA versus the rotation angles $\vartheta_x$ and $\vartheta_y$ between the transmit UCA and the receive UCA. We set the distance between transmitter and receiver is 20 m. The rotation angles $\vartheta_x$ and $\vartheta_y$ range from 0 to $\pi/3$ since there are three sectors at the transmitter and receiver. When $\vartheta_x$ and $\vartheta_y$ are large, we can observe the error is large, causing that the channel matrix cannot be regarded as a circulant matrix. Two RIS UCAs are placed in the middle of the transmitter and receiver and the rotation angles for each segment are smaller than the rotation angles $\vartheta_x$ and $\vartheta_y$.

Fig. 5 shows the error $\epsilon^t$ versus the rotation angles $\vartheta_x$ and $\vartheta_y$ for the UCA based RIS-assisted MIMO system, where we set $-\vartheta_y^{tr} = \vartheta_y^{Rr} = \vartheta_y^{rr}$, $-\vartheta_x^{tr} = \vartheta_x^{Rr} = \vartheta_x^{rr}$, $d^{tr}$= 6 m, $d^{Rr}$= 8 m, $d^{rr}$= 6 m, and $\varrho = 10^{-3}$. The distance between transmitter and receiver is also 20 m. We can obtain that when the rotation angles $\vartheta_x$ and $\vartheta_y$ are large and close to $\pi/3$, $\vartheta_y^{tr}$, $\vartheta_x^{tr}$, $\vartheta_y^{Rr}$, $\vartheta_x^{Rr}$, $\vartheta_y^{rr}$, and $\vartheta_x^{rr}$ are close to $\pi/9$ and they are within the ranges obtained based on the above $\varrho$. One can see that the error $\epsilon^t$ in Fig. 5 is much smaller than the error in Fig. 4, and we have $\widetilde{\boldsymbol{H}}_{rr}\widetilde{\boldsymbol{H}}_{Rr}\widetilde{\boldsymbol{H}}_{tr} \approx \widehat{\boldsymbol{H}}_{rr}\widehat{\boldsymbol{H}}_{Rr}\widehat{\boldsymbol{H}}_{tr}$.

**Algorithm 1** : An interpolation method for finding the ranges of rotation angles

1: $u_1 = 0$ — the minimum initial boundary of $\vartheta_x^{tr}$
$v_1 = \pi/3$ — the maximum initial boundary of $\vartheta_x^{tr}$
$u_2 = 0$ — the minimum initial boundary of $\vartheta_y^{tr}$
$v_2 = \pi/3$ — the maximum initial boundary of $\vartheta_y^{tr}$
2: Set $\varrho$
3: Set $\beta$, $\lambda$, $R_t$, $r_1$, $d^{tr}$, $\theta_t$, $\phi_t$, $\alpha_{R_t}$, and $\alpha_{r_1}$
4: $o_1 = 0$, $o_3 = 0$ — the values of $\vartheta_y^{tr}$
5: $o_2 = 0$ — the value of $\vartheta_y^{tr}$
6: $w_2 = \frac{u_1+v_1}{2}$; $w_1 = u_1$; $w_3 = v_1$ — the values of $\vartheta_x^{tr}$
7: Calculate the error $\epsilon(w_2, o_2)$ in (34).
8: **if** $\epsilon(w_2, o_2) < \varrho$ **then**
9: $v_1 = w_2$; Go to **Step 5**;
10: **else if** $\epsilon(w_2, o_2) = \varrho$ **then**
11: $o_2 = \frac{u_2+v_2}{2}$; $o_1 = u_2$; $o_3 = v_2$
12: Calculate the error $\epsilon(w_2, o_2)$ in (34).
13: **if** $\epsilon(w_2, o_2) < \varrho$ **then**
14: $v_2 = o_2$; Go to **Step 10**;
15: **else if** $\epsilon(w_2, o_2) = \varrho$ **then**
16: **if** $w_2 = o_2$ **then**
17: Get the ranges of $\vartheta_x^{tr}$ and $\vartheta_y^{tr}$;
18: **else**
19: Go to **Step 5**;
20: **end if**
21: **else if** $\epsilon(w_2, o_2) > \varrho$ **then**
22: $u_2 = o_1$; Go to **Step 10** ;
23: **end if**
24: **else if** $\epsilon(w_2, o_2) > \varrho$ **then**
25: $u_1 = w_2$; Go to **Step 5**;
26: **end if**

$$\widehat{h}_{(k+1)(n+1)} = \frac{\beta^3\lambda^3 e^{-j\frac{2\pi}{\lambda}\left(\sqrt{(d^{rr})^2+R_r^2+r_2^2}+\sqrt{(d^{Rr})^2+r_1^2+r_2^2}\sqrt{(d^{tr})^2+R_t^2+r_1^2}\right)}}{64\pi^3\sqrt{(d^{rr})^2+R_r^2+r_2^2}\sqrt{(d^{Rr})^2+r_1^2+r_2^2}\sqrt{(d^{tr})^2+R_t^2+r_1^2}} \sum_{m=1}^{N}\sum_{u=1}^{N} e^{-j\frac{2\pi}{\lambda}\left(\frac{\hat{\mathcal{D}}^{rr}_{(k+1)u}}{\sqrt{(d^{rr})^2+R_r^2+r_2^2}}+\frac{\hat{\mathcal{D}}^{Rr}_{um}}{\sqrt{(d^{Rr})^2+r_1^2+r_2^2}}+\frac{\hat{\mathcal{D}}^{tr}_{m(n+1)}}{\sqrt{(d^{tr})^2+R_t^2+r_1^2}}\right)}. \quad (38)$$

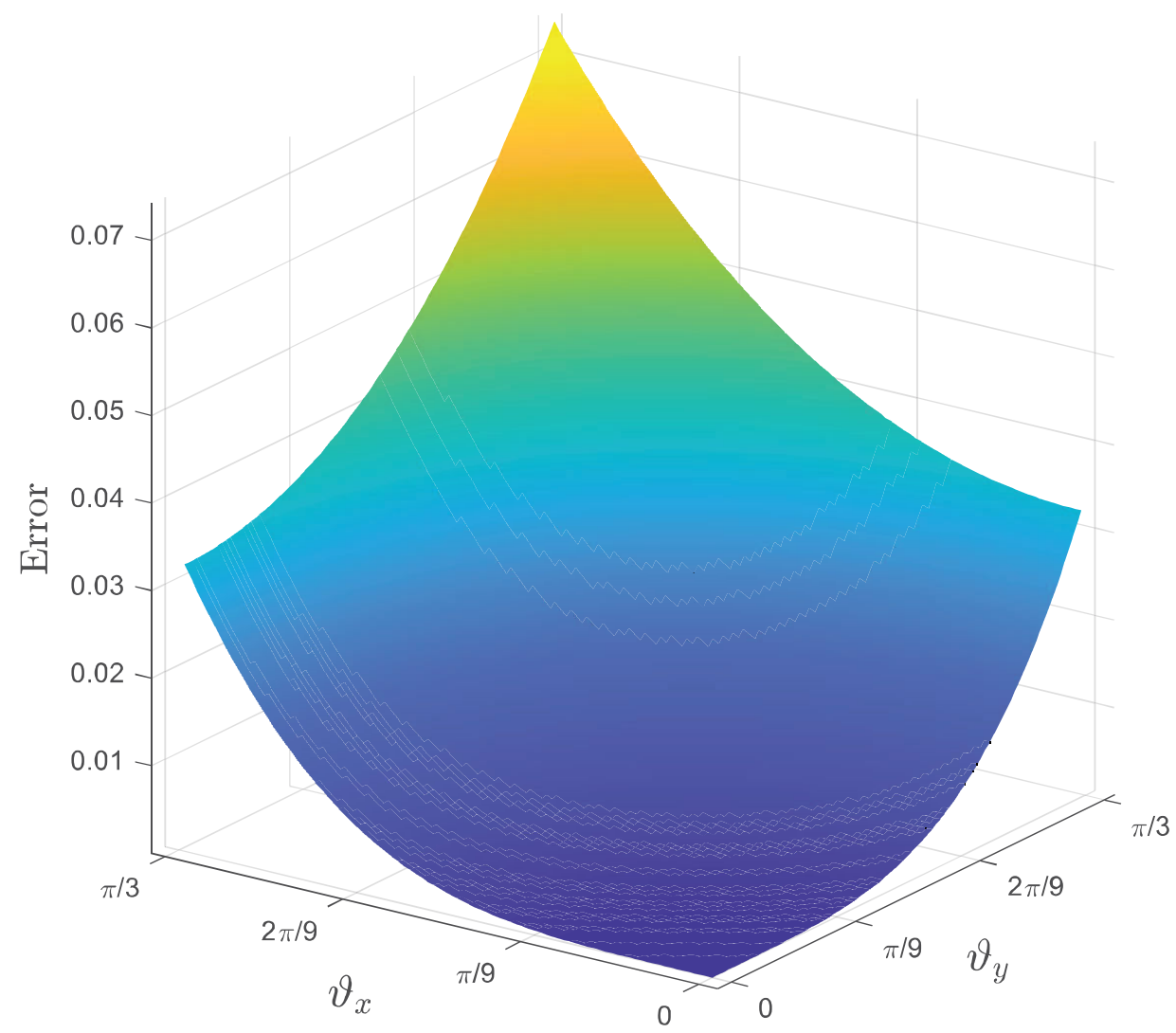

Fig. 4. The error versus the rotation angles $\vartheta_x$ and $\vartheta_y$ for the UCA based MIMO system without RISs.

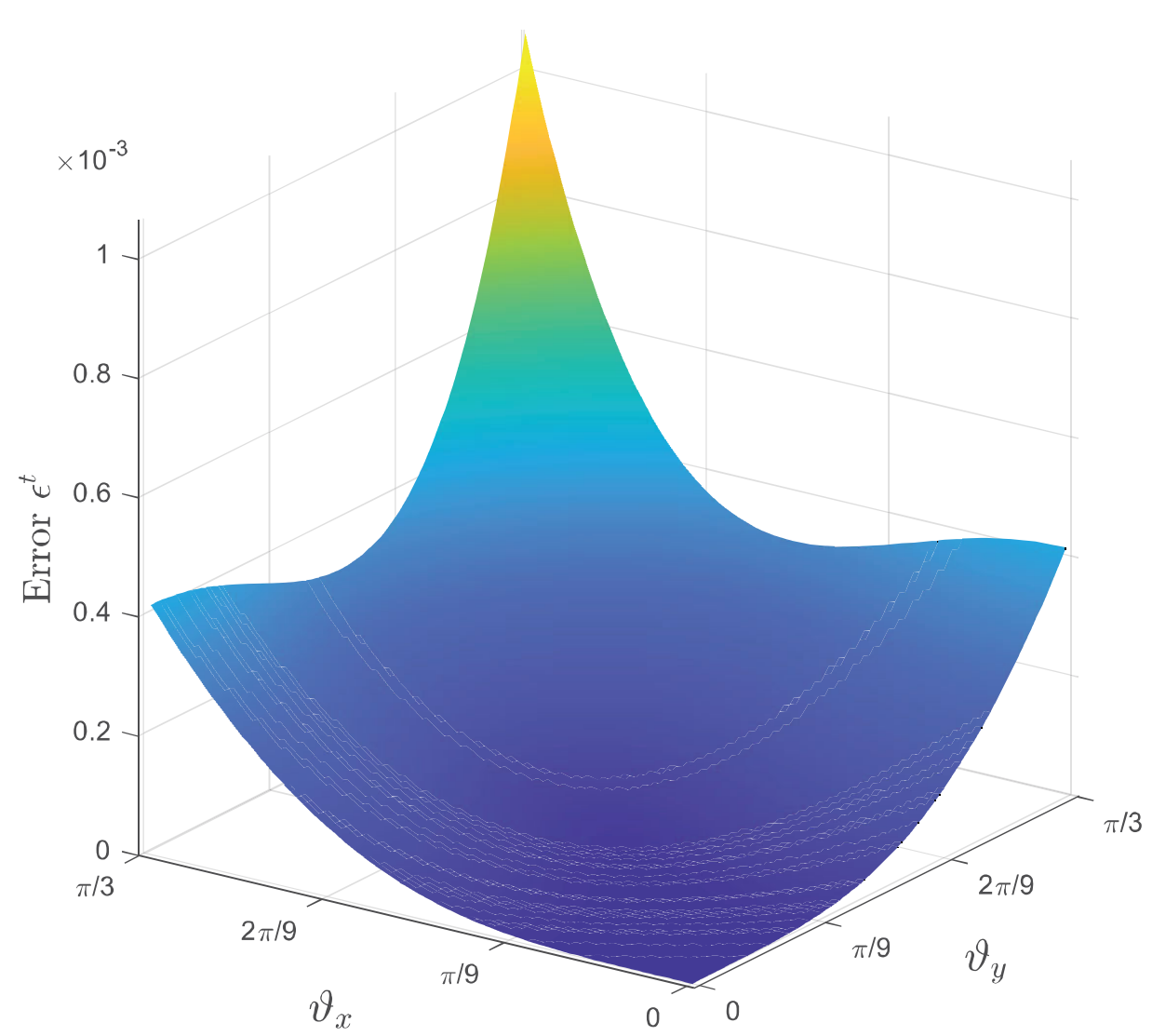

Fig. 5. The error $\epsilon^t$ versus the rotation angles $\vartheta_x$ and $\vartheta_y$ for the UCA based RIS-assisted MIMO system.

In the following, the overall channel matrix $\widehat{\boldsymbol{H}}_{rr}\widehat{\boldsymbol{H}}_{Rr}\widehat{\boldsymbol{H}}_{tr}$ is proved to be a circulant matrix. Denote by $\widehat{h}_{kn}$, derived in (37), the $k$th row and $n$th column element of $\widehat{\boldsymbol{H}}_{rr}\widehat{\boldsymbol{H}}_{Rr}\widehat{\boldsymbol{H}}_{tr}$. Then, we derive $\widehat{h}_{(k+1)(n+1)}$, $1 \le k, n \le N-1$, given in (38).

Since

$$\begin{aligned}
&\hat{\mathcal{D}}^{rr}_{(k+1)u}\\
&=\frac{R_r r_2}{-2}(1+\cos\vartheta_y^{rr}\cos\vartheta_x^{rr})\cos\left(\psi_k-\varphi_u+\alpha_{R_r}-\alpha_{r2}-\chi_1^{rr}+\frac{2\pi}{N}\right)\\
&=\frac{R_r r_2}{-2}(1+\cos\vartheta_y^{rr}\cos\vartheta_x^{rr})\Bigg[\cos\left(\psi_k-\varphi_u+\alpha_{R_r}-\alpha_{r2}-\chi_1^{rr}\right)-\\
&\sqrt{2-2\cos\frac{2\pi}{N}}\cos\left(\psi_k-\varphi_u+\alpha_{R_r}-\alpha_{r2}-\chi_1^{rr}-\arctan\frac{\sin\frac{2\pi}{N}}{1-\cos\frac{2\pi}{N}}\right)\Bigg]\\
&=\hat{\mathcal{D}}^{rr}_{k1}+A^r\sqrt{(d^{rr})^2+R_r^2+r_2^2}\\
&\times\cos\left(\psi_k-\varphi_u+\alpha_{R_r}-\alpha_{r2}-\chi_1^{rr}-\arctan\frac{\sin(2\pi/N)}{1-\cos(2\pi/N)}\right) \quad (39)
\end{aligned}$$

and

$$\begin{aligned}
&\hat{\mathcal{D}}^{tr}_{m(n+1)}\\
&=\frac{R_t r_1}{-2}(1+\cos\vartheta_y^{tr}\cos\vartheta_x^{tr})\cos\left(\psi_n-\varphi_m+\alpha_{R_t}-\alpha_{r1}-\chi_1^{tr}+\frac{2\pi}{N}\right)\\
&=\frac{R_t r_q}{-2}(1+\cos\vartheta_y^{tr}\cos\vartheta_x^{tr})\Bigg[\cos\left(\psi_n-\varphi_m+\alpha_{R_t}-\alpha_{r1}-\chi_1^{tr}\right)-\\
&\sqrt{2-2\cos\frac{2\pi}{N}}\cos\left(\psi_n-\varphi_m+\alpha_{R_t}-\alpha_{r1}-\chi_1^{tr}-\arctan\frac{\sin\frac{2\pi}{N}}{1-\cos\frac{2\pi}{N}}\right)\Bigg]\\
&=\hat{\mathcal{D}}^{tr}_{mn}+A^r\sqrt{(d^{tr})^2+R_t^2+r_1^2}\\
&\times\cos\left(\psi_n-\varphi_m+\alpha_{R_t}-\alpha_{r1}-\chi_1^{tr}-\arctan\frac{\sin(2\pi/N)}{1-\cos(2\pi/N)}\right), \quad (40)
\end{aligned}$$

$\widehat{h}_{(k+1)(n+1)}$ can be derived in Eq. (41) at the next page. We rewrite some items in (41) as $A^t$ and $A^r$ for convenience.

We can observe that the factors in front of the summation signs in (37) and (41) are the same. The following Lemma 2 demonstrates that the phases of the terms in the summations in (37) and (41) are the same as well.

**Lemma 2.**

$$\begin{aligned}
&\sum_{m=1}^{N}\sum_{u=1}^{N}e^{-j\frac{2\pi}{\lambda}\left(\frac{\hat{\mathcal{D}}^{rr}_{ku}}{\sqrt{(d^{rr})^2+R_r^2+r_2^2}}+\frac{\hat{\mathcal{D}}^{Rr}_{um}}{\sqrt{(d^{Rr})^2+r_1^2+r_2^2}}+\frac{\hat{\mathcal{D}}^{tr}_{mn}}{\sqrt{(d^{tr})^2+R_t^2+r_1^2}}\right)}\\
&=\sum_{m=1}^{N}\sum_{u=1}^{N}\Bigg[e^{-j\frac{2\pi}{\lambda}\left(\frac{\hat{\mathcal{D}}^{rr}_{ku}}{\sqrt{(d^{rr})^2+R_r^2+r_2^2}}+\frac{\hat{\mathcal{D}}^{Rr}_{um}}{\sqrt{(d^{Rr})^2+r_1^2+r_2^2}}+\frac{\hat{\mathcal{D}}^{tr}_{mn}}{\sqrt{(d^{tr})^2+R_t^2+r_1^2}}\right)}\\
&\times e^{-j\frac{2\pi}{\lambda}A^t\cos\left(\psi_n-\varphi_m+\alpha_{R_t}-\alpha_{r1}-\chi_1^{tr}-\arctan\frac{\sin\frac{2\pi}{N}}{1-\cos\frac{2\pi}{N}}\right)}\\
&\times e^{-j\frac{2\pi}{\lambda}A^r\cos\left(\psi_k-\varphi_u+\alpha_{R_r}-\alpha_{r2}-\chi_1^{rr}-\arctan\frac{\sin\frac{2\pi}{N}}{1-\cos\frac{2\pi}{N}}\right)}\Bigg] \quad (42)
\end{aligned}$$

*when $k$ and $n$ range from 1 to $N-1$.*

$$\widehat{h}_{(k+1)(n+1)} =$$

$$\frac{\beta^3\lambda^3 e^{-j\frac{2\pi}{\lambda}\left(\sqrt{(d^{rr})^2+R_r^2+r_2^2}+\sqrt{(d^{Rr})^2+r_1^2+r_2^2}\sqrt{(d^{tr})^2+R_t^2+r_1^2}\right)}}{64\pi^3\sqrt{(d^{rr})^2+R_r^2+r_2^2}\sqrt{(d^{Rr})^2+r_1^2+r_2^2}\sqrt{(d^{tr})^2+R_t^2+r_1^2}}\sum_{m=1}^{N}\sum_{u=1}^{N}\left\{e^{-j\frac{2\pi}{\lambda}\left(\frac{\hat{\mathcal{D}}_{ku}^{rr}}{\sqrt{(d^{rr})^2+R_r^2+r_2^2}}+\frac{\hat{\mathcal{D}}_{um}^{Rr}}{\sqrt{(d^{Rr})^2+r_1^2+r_2^2}}+\frac{\hat{\mathcal{D}}_{mn}^{tr}}{\sqrt{(d^{tr})^2+R_t^2+r_1^2}}\right)}\right.$$

$$\times\exp\left[-j\frac{2\pi}{\lambda}\underbrace{\frac{\sqrt{1-\cos(2\pi/N)}R_t r_1(1+\cos\vartheta_y^{tr}\cos\vartheta_x^{tr})}{\sqrt{2}\sqrt{(d^{tr})^2+R_t^2+r_1^2}}}_{A^t}\cos\left(\psi_n-\varphi_m+\alpha_{R_t}-\alpha_{r1}-\chi_1^{tr}-\arctan\frac{\sin 2\pi/N}{1-\cos 2\pi/N}\right)\right]$$

$$\left.\times\exp\left[-j\frac{2\pi}{\lambda}\underbrace{\frac{\sqrt{1-\cos(2\pi/N)}R_r r_2(1+\cos\vartheta_y^{rr}\cos\vartheta_x^{rr})}{\sqrt{2}\sqrt{(d^{rr})^2+R_r^2+r_2^2}}}_{A^r}\cos\left(\psi_k-\varphi_u+\alpha_{R_r}-\alpha_{r2}-\chi_1^{rr}-\arctan\frac{\sin 2\pi/N}{1-\cos 2\pi/N}\right)\right]\right\}. \quad (41)$$

*Proof.* See Appendix B. □

According to Lemma 2, we can obtain that $\widehat{h}_{kn} = \widehat{h}_{(k+1)(n+1)}$, $1 \le k, n \le N-1$. Similarly, when $k = N$ and $1 \le n \le N-1$, we can obtain $\widehat{h}_{Nn} = \widehat{h}_{1(n+1)}$, and when $1 \le k \le N-1$ and $n = N$, we can obtain $\widehat{h}_{kN} = \widehat{h}_{(k+1)1}$. Therefore, the overall channel matrix $\widehat{\boldsymbol{H}}_{rr}\widehat{\boldsymbol{H}}_{Rr}\widehat{\boldsymbol{H}}_{tr}$ is a circulant matrix.

### E. Multiple RISs Placed Between Transmitter and Receiver

When there are more than two RISs placed between the transmitter and the receiver, similar to the scenario with two RIS UCAs considered above, the phase compensation matrices on RIS UCAs can be derived based on the rotation angles for each segment. Then, the overall channel matrix also can be approximated into a circulant matrix after phase compensations on RIS UCAs and the proof can be obtained using Lemma 2. Observing Fig. 3, we can obtain that the error decreases as the rotation angles decrease and the error is close to zero when the rotation angles are small enough. The rotation angles on each segment become smaller when more RISs are placed in the middle of the transmitter and the receiver. Thus, the error between the overall channel model and the overall approximated channel model from the transmit UCA to the receive UCA is small when there are multiple RISs.

## IV. Achieving Fast Detection For UCA Based RIS-Assisted MIMO System

In this section, we introduce the precoding scheme on the transmitter, the phase designs on the RIS UCAs, as well as the phase compensation scheme and the fast detection scheme on the receiver.

### A. Precoding Scheme on the Transmitter

We propose a precoding scheme where signals at different antenna elements have different phases. The transmit signal of the $n$th, $1 \le n \le N$, antenna element on the transmit UCA, denoted by $x_n$, is given as follows:

$$x_n = \sum_{l=1}^{N}\frac{1}{\sqrt{N}}s_l e^{j(\psi_n+\alpha_{R_t})(l-1)}e^{j\frac{2\pi}{\lambda}\frac{\varepsilon_n^{tr}}{\sqrt{(d^{tr})^2+R_t^2+r_1^2}}}, \quad (43)$$

where $s_1$, $s_2$, $\ldots$, $s_N$ represent the $N$ information symbols to send.

Then, the vector corresponding to the transmit signals, denoted by $\boldsymbol{x}$, in (43) can be written as:

$$\boldsymbol{x} = \boldsymbol{U}_1^*\boldsymbol{W}\boldsymbol{s}, \quad (44)$$

where $\boldsymbol{s} = [s_1, s_2, \cdots, s_N]^T$ and $\boldsymbol{W}$ is the $N$-point inverse discrete Fourier transform (IDFT) matrix, $(\cdot)^*$ represents the conjugation operation, and $\boldsymbol{U}_1$ is a unitary and diagonal matrix given in (15). The precoding matrix is $\boldsymbol{U}_1^*\boldsymbol{W}$ and the signals after IDFT are multiplied by the $N$ diagonal elements of the unitary diagonal matrix $\boldsymbol{U}_1^*$. Thus, for the precoding scheme, the number of complex additions is $N\log_2(N)$ and the number of complex multiplications is $N/2\log_2(N)+N$.

### B. Phase Designs on the RIS UCAs

RIS is a planer surface comprising reflecting elements and each element is able to introduce phase change to the incident signals independently using the varactors on the elements. The incident signals are reflected directly after phase changes without demodulation and amplification. For generality, the noise on the RIS is taken into consideration in this paper. We give the phase design on the RIS UCAs to change the phases of the incident signals and help the receive UCA to demodulate signals with fast schemes. The incident signal vector, denote by $\boldsymbol{y}_1^r$, on the RIS-1 UCA is

$$\boldsymbol{y}_1^r = \boldsymbol{H}_{tr}\boldsymbol{x} + \boldsymbol{n}_{tr}, \quad (45)$$

where $\boldsymbol{n}_{tr}$ is the noise vector on the RIS-1 UCA and has mean 0 and variance $\sigma_1^2$. Since each element on the RIS-1 UCA can change the phase corresponding to the incident signals independently, the phase compensation matrix, denoted by $\boldsymbol{H}_1^r$, on the RIS-1 UCA is

$$\boldsymbol{H}_1^r = \boldsymbol{U}_3^*\boldsymbol{U}_2^*, \quad (46)$$

where $\boldsymbol{U}_2^*$ and $\boldsymbol{U}_3^*$ are the phase matrices on the RIS-1 UCA to compensate the partial phases of $\boldsymbol{H}_{tr}$ and $\boldsymbol{H}_{Rr}$ defined in (15) and (23), respectively. From (15) and (23), both $\boldsymbol{U}_2$ and $\boldsymbol{U}_3$ are diagonal and unitary. Thus, $\boldsymbol{H}_1^r$ is also diagonal and unitary. The reflected signal vector, denoted by $\boldsymbol{x}_1^r$, on the RIS-1 UCA is

$$\boldsymbol{x}_1^r = \boldsymbol{H}_1^r \boldsymbol{y}_1^r. \tag{47}$$

The compensation phase is applied to the varactor on each element of RIS-1 UCA to change the phases of the incident signals by adjusting the bias voltage. The number of complex additions is 0 and the number of phase subtractions is $N$. Then, the incident signal vector, denote by $\boldsymbol{y}_2^r$, on the RIS-2 UCA is

$$\boldsymbol{y}_2^r = \boldsymbol{H}_{Rr}\boldsymbol{x}_1^r + \boldsymbol{n}_{Rr}, \tag{48}$$

where $\boldsymbol{n}_{Rr}$ is the noise vector corresponding to the RIS-2 UCA and has mean 0 and variance $\sigma_2^2$. Let $\boldsymbol{H}_2^r$ be the phase compensation matrix on the RIS-2 UCA as:

$$\boldsymbol{H}_2^r = \boldsymbol{U}_5^* \boldsymbol{U}_4^*, \tag{49}$$

where $\boldsymbol{U}_4^*$ and $\boldsymbol{U}_5^*$ are the phase matrices on the RIS-2 UCA to compensate the partial phases $\boldsymbol{H}_{Rr}$ and $\boldsymbol{H}_{rr}$ defined in (23) and (31), respectively. From (23) and (31), both $\boldsymbol{U}_4$ and $\boldsymbol{U}_5$ are diagonal and unitary. Thus, $\boldsymbol{H}_2^r$ is also diagonal and unitary. The reflected signal vector, denoted by $\boldsymbol{x}_2^r$, on the RIS-2 UCA is

$$\boldsymbol{x}_2^r = \boldsymbol{H}_2^r \boldsymbol{y}_2^r. \tag{50}$$

Also, the compensation phases corresponding to $\boldsymbol{H}_2^r$ can be attached to the incident signals by adjusting the bias voltage applied to the varactor. The number of complex additions is 0 and the number of phase subtractions is $N$.

### *C. A Phase Compensation Scheme and Symbol-Wise ML Detection Scheme on the Receive UCA*

The receive signal vector, denoted by $\boldsymbol{y}$, on the receive UCA is

$$\begin{aligned}\boldsymbol{y} &= \boldsymbol{H}_{rr}\boldsymbol{x}_2^r + \boldsymbol{n}_{rr}\\ &= \boldsymbol{H}_{rr}\boldsymbol{H}_2^r\boldsymbol{H}_{Rr}\boldsymbol{H}_1^r\boldsymbol{H}_{tr}\boldsymbol{x}\\ &\quad + \boldsymbol{H}_{rr}\boldsymbol{H}_2^r\boldsymbol{H}_{Rr}\boldsymbol{H}_1^r\boldsymbol{n}_{tr} + \boldsymbol{H}_{rr}\boldsymbol{H}_2^r\boldsymbol{n}_{Rr} + \boldsymbol{n}_{rr},\end{aligned} \tag{51}$$

where $\boldsymbol{H}_{tr}$, $\boldsymbol{H}_{Rr}$, and $\boldsymbol{H}_{rr}$ are given in (14), (22), and (30), respectively and $\boldsymbol{n}_{rr}$ is the noise vector on the receive UCA of mean 0 and variance $\sigma_3^2$. A phase compensation is first proposed to eliminate the phase impact and recover the orthogonality among different signals. The phase compensation matrix on the receive UCA is $\boldsymbol{U}_6^*$ defined in (31) that is also diagonal and unitary. Multiplying $\boldsymbol{U}_6^*$ to the receive signals at the receive UCA, we have

$$\begin{aligned}&\boldsymbol{U}_6^*\boldsymbol{y}\\ &= \boldsymbol{U}_6^*\boldsymbol{H}_{rr}\boldsymbol{H}_2^r\boldsymbol{H}_{Rr}\boldsymbol{H}_1^r\boldsymbol{H}_{tr}\boldsymbol{x}\\ &+\boldsymbol{U}_6^*\boldsymbol{H}_{rr}\boldsymbol{H}_2^r\boldsymbol{H}_{Rr}\boldsymbol{H}_1^r\boldsymbol{n}_{tr} + \boldsymbol{U}_6^*\boldsymbol{H}_{rr}\boldsymbol{H}_2^r\boldsymbol{n}_{Rr} + \boldsymbol{U}_6^*\boldsymbol{n}_{rr}\\ &\approx \widehat{\boldsymbol{H}}_{rr}\widehat{\boldsymbol{H}}_{Rr}\widehat{\boldsymbol{H}}_{tr}\boldsymbol{W}\boldsymbol{s}\\ &\quad + \widehat{\boldsymbol{H}}_{rr}\widehat{\boldsymbol{H}}_{Rr}\boldsymbol{U}_2^*\boldsymbol{n}_{tr} + \widehat{\boldsymbol{H}}_{rr}\boldsymbol{U}_4^*\boldsymbol{n}_{Rr} + \boldsymbol{U}_6^*\boldsymbol{n}_{rr}.\end{aligned} \tag{52}$$

Since $\widehat{\boldsymbol{H}}_{rr}\widehat{\boldsymbol{H}}_{tr}\widehat{\boldsymbol{H}}_{tr}$ is a circulant matrix as previously studied, we have

$$\boldsymbol{W}^*\widehat{\boldsymbol{H}}_{rr}\widehat{\boldsymbol{H}}_{Rr}\widehat{\boldsymbol{H}}_{tr}\boldsymbol{W} = \boldsymbol{\Lambda}, \tag{53}$$

where $\boldsymbol{W}^*$ is the $N$-point DFT matrix and $\boldsymbol{\Lambda}$ is a diagonal matrix. The $i$th diagonal element of $\boldsymbol{\Lambda}$, denoted by $\Lambda_i$, $1 \leq i \leq N$, is

$$\Lambda_i = \sum_{n=1}^{N} \frac{1}{\sqrt{N}} \widehat{h}_{1n} e^{-j\frac{2\pi}{N}(i-1)(n-1)}, \tag{54}$$

where $\widehat{h}_{1n}$ is the $n$th element of the first row vector of matrix $\widehat{\boldsymbol{H}}_{rr}\widehat{\boldsymbol{H}}_{Rr}\widehat{\boldsymbol{H}}_{tr}$. Denote $\boldsymbol{W}^*\boldsymbol{U}_6^*\boldsymbol{y} = \widetilde{\boldsymbol{y}} = [\widetilde{y}_1, \widetilde{y}_2, \cdots, \widetilde{y}_N]^T$. Then, the ML estimate of the signal vector $\boldsymbol{s}$ is

$$\begin{aligned}\widehat{\boldsymbol{s}} &= \arg\min_{\boldsymbol{s}\in\boldsymbol{\Omega}^N} \|\widetilde{\boldsymbol{y}} - \boldsymbol{\Lambda}\boldsymbol{s}\|^2\\ &= \arg\min_{\boldsymbol{s}\in\boldsymbol{\Omega}^N} \sum_{l=1}^{N} |\widetilde{y}_l - \Lambda_l s_l|^2\\ &= \left[\arg\min_{s_1\in\boldsymbol{\Omega}} |\widetilde{y}_1 - \Lambda_1 s_1|^2, \cdots, \arg\min_{s_N\in\boldsymbol{\Omega}} |\widetilde{y}_N - \Lambda_N s_N|^2\right]^T\\ &= \left[\arg\min_{s_1\in\boldsymbol{\Omega}} |\widetilde{y}_1 - \Lambda_1 s_1|, \cdots, \arg\min_{s_N\in\boldsymbol{\Omega}} |\widetilde{y}_N - \Lambda_N s_N|\right]^T,\end{aligned} \tag{55}$$

where $\boldsymbol{\Omega}$ is a signal constellation of size $V$.

For the traditional RIS-assisted MIMO system where there are no precoding scheme on the transmitter, no phase designs on the RIS UCAs, and no phase compensation scheme on the receiver, each antenna on the transmitter emits one symbol and RIS UCAs only reflect the incident signals. Then, the corresponding ML detection is

$$\widehat{\boldsymbol{s}} = \arg\min_{\boldsymbol{s}\in\boldsymbol{\Omega}^N} \|\boldsymbol{y} - \boldsymbol{H}_{rr}\boldsymbol{H}_{Rr}\boldsymbol{H}_{tr}\boldsymbol{s}\|^2. \tag{56}$$

## V. BER, Capacity, and Computational Complexity Analyses

For the UCA based RIS-assisted MIMO system, where the transmit UCA, the RIS UCAs, and the receive UCA are not aligned with each other, the BER, denoted by $P_{eg}$, corresponding to binary phase shift keying (BPSK), using our proposed schemes can be derived as follows:

$$P_{eg} = \frac{1}{N}\sum_{l=1}^{N}\frac{1}{2}\mathrm{erfc}\left(\frac{|\Lambda_l|^2|s_l|^2}{N\omega_l^2}\right), \tag{57}$$

where $\mathrm{erfc}(x) = \frac{2}{\sqrt{\pi}}\int_x^{\infty} e^{-t^2}dt$ and $\omega_l^2$ represents the variance of the $l$th received noise in (52) for the UCA based RIS-assisted MIMO system with misaligned transceiver. Assume the noises on RIS-1 UCA, RIS-2 UCA, and the receive UCA are independent. Since unitary transforms do not change the noise properties, from (52) $\omega_l^2$ is derived as:

$$\begin{aligned}\omega_l^2 &= \frac{1}{N}\left|\sum_{u=1}^{N}\sum_{m=1}^{N}\widehat{h}_{1u}^{rr}\widehat{h}_{1m}^{Rr}e^{-j\frac{2\pi}{N}(l-1)(u+m-2)}\right|^2\sigma_1^2\\ &\quad + \frac{1}{\sqrt{N}}\left|\sum_{u=1}^{N}\widehat{h}_{1u}^{rr}e^{-j\frac{2\pi}{N}(l-1)(u-1)}\right|^2\sigma_2^2 + \sigma_3^2.\end{aligned} \tag{58}$$

The capacity, denoted by $C_{eg}$, of the UCA based RIS-assisted MIMO system using our proposed scheme is

$$C_{eg} = B\sum_{l=1}^{N}\log\left(1 + \frac{|\Lambda_l|^2|s_l|^2}{N\omega_l^2}\right), \tag{59}$$

where $B$ represents the bandwidth. When $N$ is large, the channels are correlated and the overall channel matrix may not be full-rank.

TABLE I
COMPLEXITY COMPARISON AT THE TRANSMITTER FOR THE MISALIGNED SCENARIO

| Precoding scheme | The number of complex additions | The number of complex multiplications |
|---|---|---|
| Precoding scheme | $N\log_2(N)$ | $\frac{N}{2}\log_2(N)+N$ |
| No precoding scheme | 0 | 0 |
| Fully digital precoding scheme | $\frac{4}{3}N^3-\frac{3}{2}N^2+\frac{1}{6}N$ | $\frac{5}{3}N^3+\frac{1}{2}N^2-\frac{1}{6}N$ |

In view of the computational complexity, the number of complex additions is $N\log_2(N)$ and the number of complex multiplications is $N/2\log_2(N)+N$ for the precoding scheme on the transmitter. Since RISs only reflect the signals and do not demodulate the signals, the number of complex additions is 0 and the number of phase subtractions is $N$ on each RIS UCA. Using our proposed fast symbol-wise detection scheme, the number of complex additions is $N\log_2(N)+NV$ and the number of complex multiplications is $N/2\log_2(N)+N(V+1)$ on the receiver, where $V$ represents the size of modulation alphabet.

TABLE II
COMPLEXITY COMPARISON AT THE RECEIVER FOR THE MISALIGNED SCENARIO

| Detection scheme | The number of complex additions | The number of complex multiplications |
|---|---|---|
| Fast symbol-wise ML | $N\log_2(N)+NV$ | $\frac{N}{2}\log_2(N)+N(V+1)$ |
| Traditional ML | $N^2V^N$ | $(N^2+N)V^N$ |
| ML with fully digital precoding | $NV$ | $NV$ |

TABLE III
COMPUTATIONAL COMPLEXITIES AT RIS-1 UCA AND RIS-2 UCA

| Phase compensation Scheme | The number of complex additions | The number of phase subtractions |
|---|---|---|
| RIS-1 UCA | 0 | $N$ |
| RIS-2 UCA | 0 | $N$ |

With the misaligned transceiver, the computational complexities at the transmitter corresponding to our proposed precoding scheme, no precoding scheme, and the fully digital precoding scheme are listed in Table I. By fully digital precoding at the transmitter, it means that the digital precoding completely removes the overall channel matrix. The computational complexities at the receiver corresponding to our proposed fast symbol-wise ML detection scheme, the traditional ML detection scheme, and ML detection scheme with the fully digital precoding scheme, where the overall channel is assumed known at the transmitter and fully compensated by the precoding, are listed in Table II. The computational complexities at the two RIS UCAs are listed in Table III. Only the phases of incident signals are changed according to the $N$ diagonal elements of the unitary diagonal phase compensation matrices on the RIS UCAs. For example, when $N = 20$ and $V = 4$, the numbers of additions and multiplications for the fully digital precoding scheme are approximately 115 and 213 times, for the traditional ML scheme are approximately $2.64\times10^{12}$ and $3.22\times10^{12}$ times more than the numbers of additions and multiplications for our proposed fast symbol-wise ML scheme on the receiver, respectively.

## VI. SIMULATION RESULTS

In this section, we use simulations to evaluate our proposed precoding scheme on the transmitter, the phase designs on the RIS UCAs, the phase compensation scheme on the receiver, and the fast symbol-wise ML detection scheme for the UCA based RIS-assisted MIMO system with the misaligned transceiver. We compare the BERs and capacities of our proposed transceiver design and the traditional ML detection (56). The simulation parameters are given in Table IV.

TABLE IV
SIMULATION PARAMETERS

| | | | |
|---|---|---|---|
| $\beta$ | $4\pi$ | $\lambda$ | 0.003 m |
| $R_t$ | 0.12 m | $R_r$ | 0.12 m |
| $r_1$ | 0.12 m | $r_2$ | 0.12 m |
| $d^{tr}$ | 4.2 m | $d^{Rr}$ | 4.5 m |
| $d^{rr}$ | 4.2 m | B | 1 MHz |
| $\alpha_{R_t}$ | 0 | $\alpha_{R_r}$ | 0 |
| $\alpha_{r_1}$ | 0 | $\alpha_{r_2}$ | 0 |

Fig. 6 depicts the BERs of the UCA based RIS-assisted MIMO system using our proposed transceiver design and using the traditional ML detection for the misaligned transceiver scenario. We can observe that the BERs of the UCA based RIS-assisted MIMO system using our proposed transceiver design and using the traditional ML detection are the same. This is because the precoding matrix on the transmitter, the phase compensation matrices on the RIS UCAs, and the phase compensation matrix on the receiver are unitary. Thus, they do not change the channel properties. Also, the BERs increase as the number $N$ of antennas increases for the UCA based RIS-assisted MIMO system using our proposed transceiver design and using the traditional ML detection. This is because the number $N$ of the information symbols per channel use to send increases as well.

Fig. 7 shows the BERs versus included angles corresponding to the UCA based RIS-assisted MIMO system using our proposed transceiver design and using the traditional ML detection for the misaligned transceiver. We can observe that the included angles have no impact on the BER. This is because the matrices with the elements including the included angles are unitary and they do not change the channel properties for the UCA based RIS-assisted MIMO system. The BERs for the UCA based RIS-assisted MIMO system using our proposed transceiver design and using the traditional ML detection are the same. This is because the precoding matrix on the transmitter, the phase design matrices on the RIS UCAs, and the phase compensation matrix on the receiver are unitary.

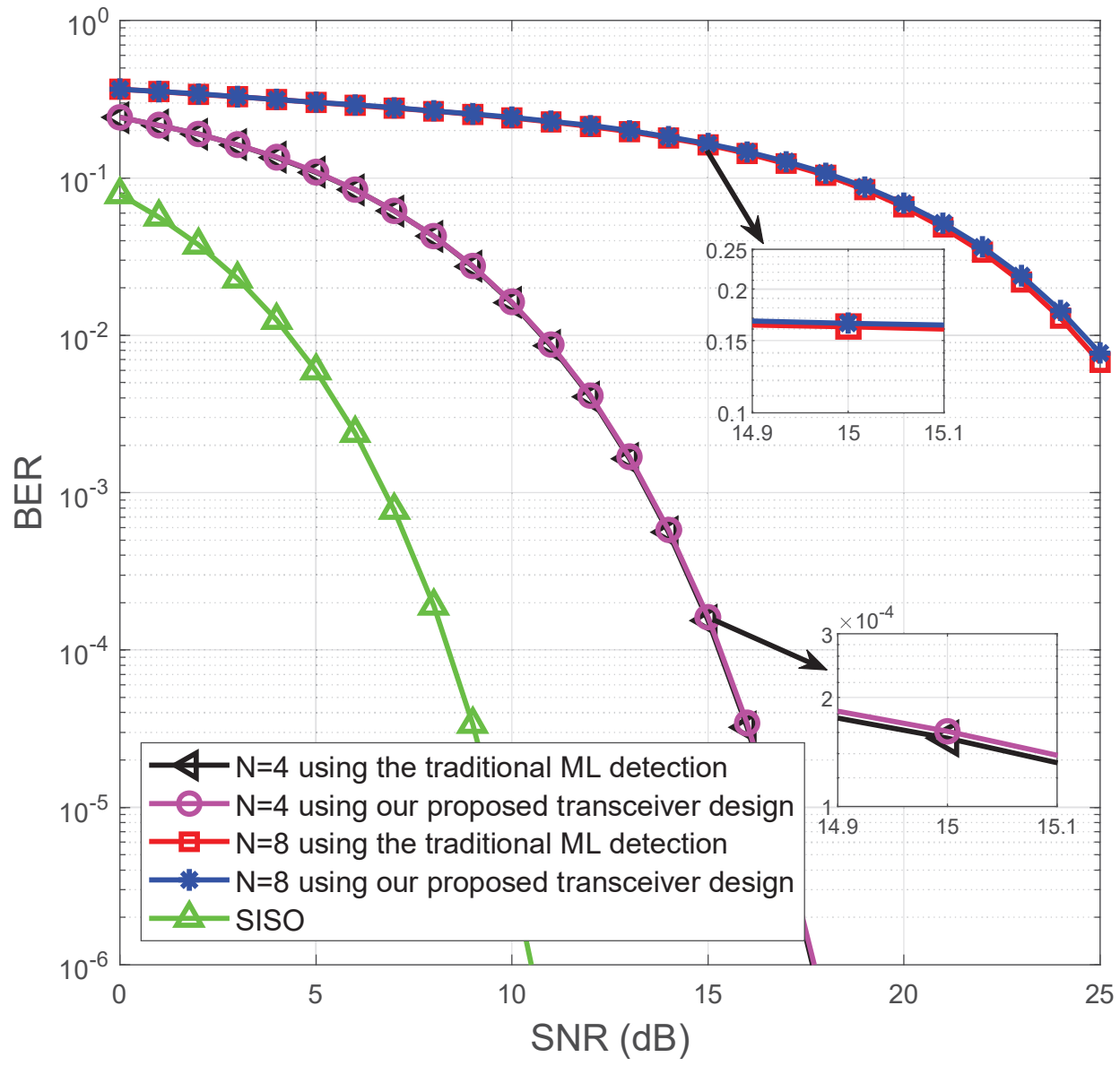

Fig. 6. BERs for UCA based RIS-assisted MIMO system using our proposed transceiver design and using the traditional ML detection.

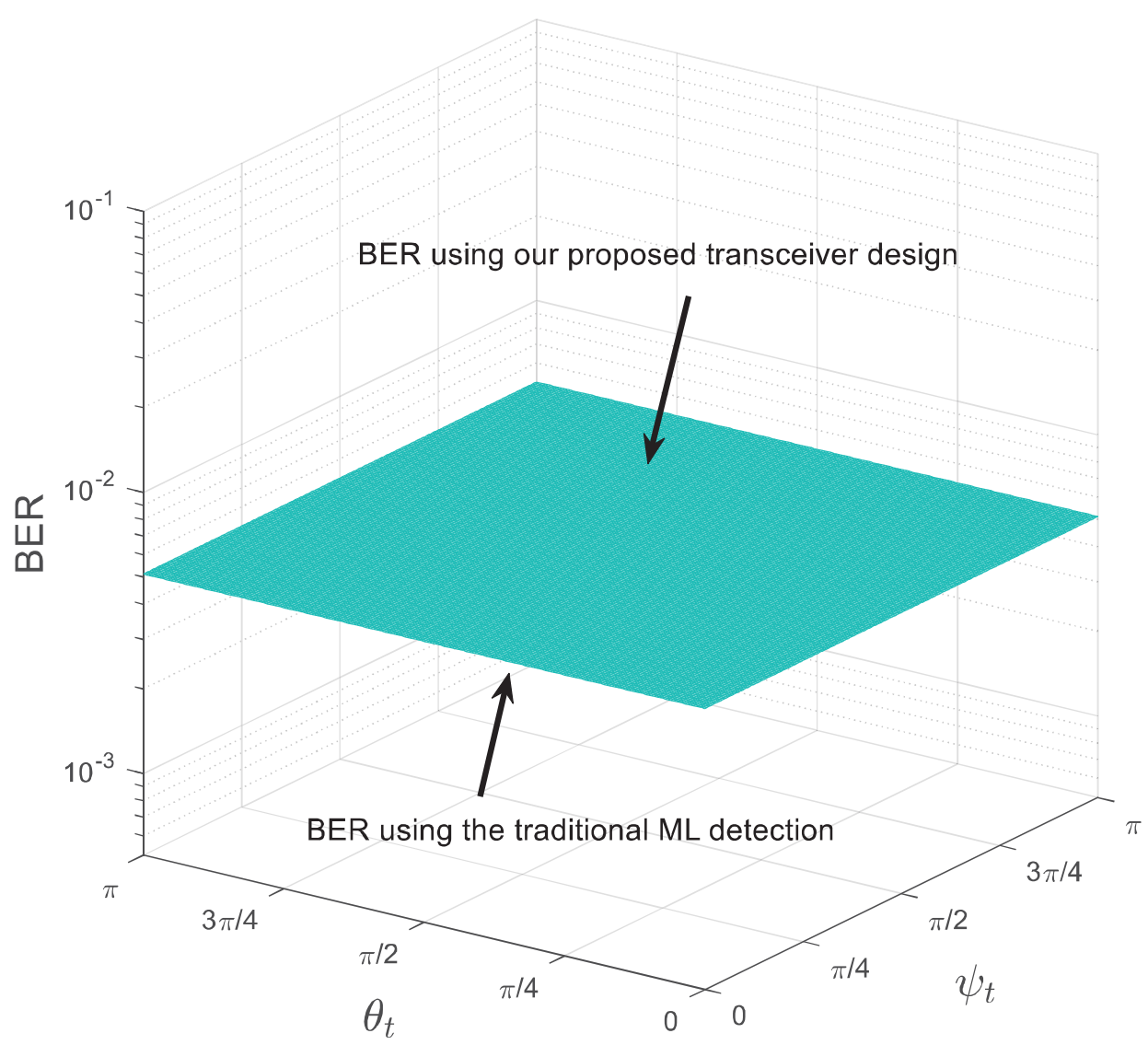

Fig. 7. BERs versus included angles for the UCA based RIS-assisted MIMO system where the transmitter, the RIS UCAs, and the receiver are not aligned.

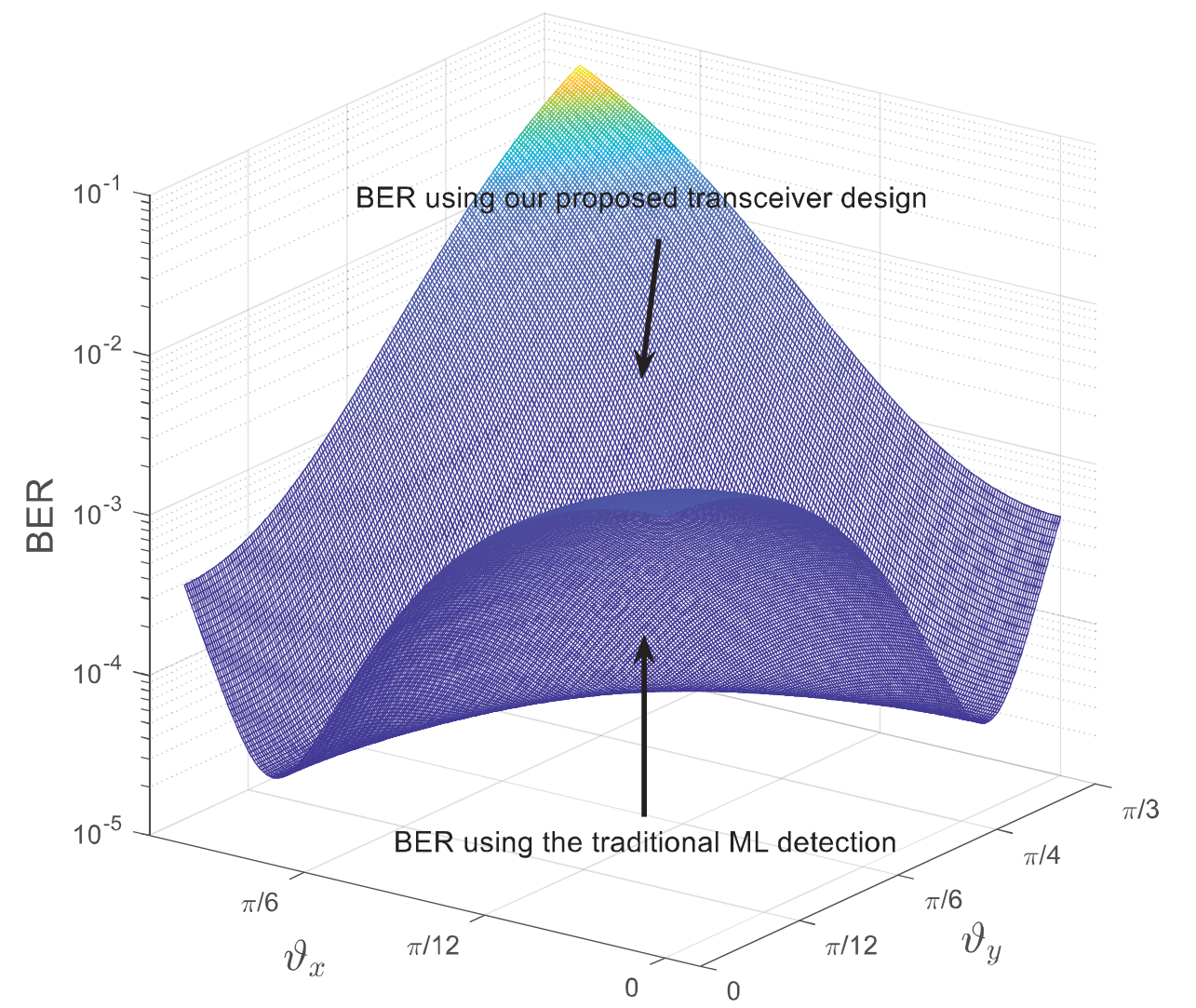

Fig. 8. BERs versus rotation angles for the UCA based RIS-assisted MIMO system where the transmitter, the RIS UCAs, and the receiver are not aligned.

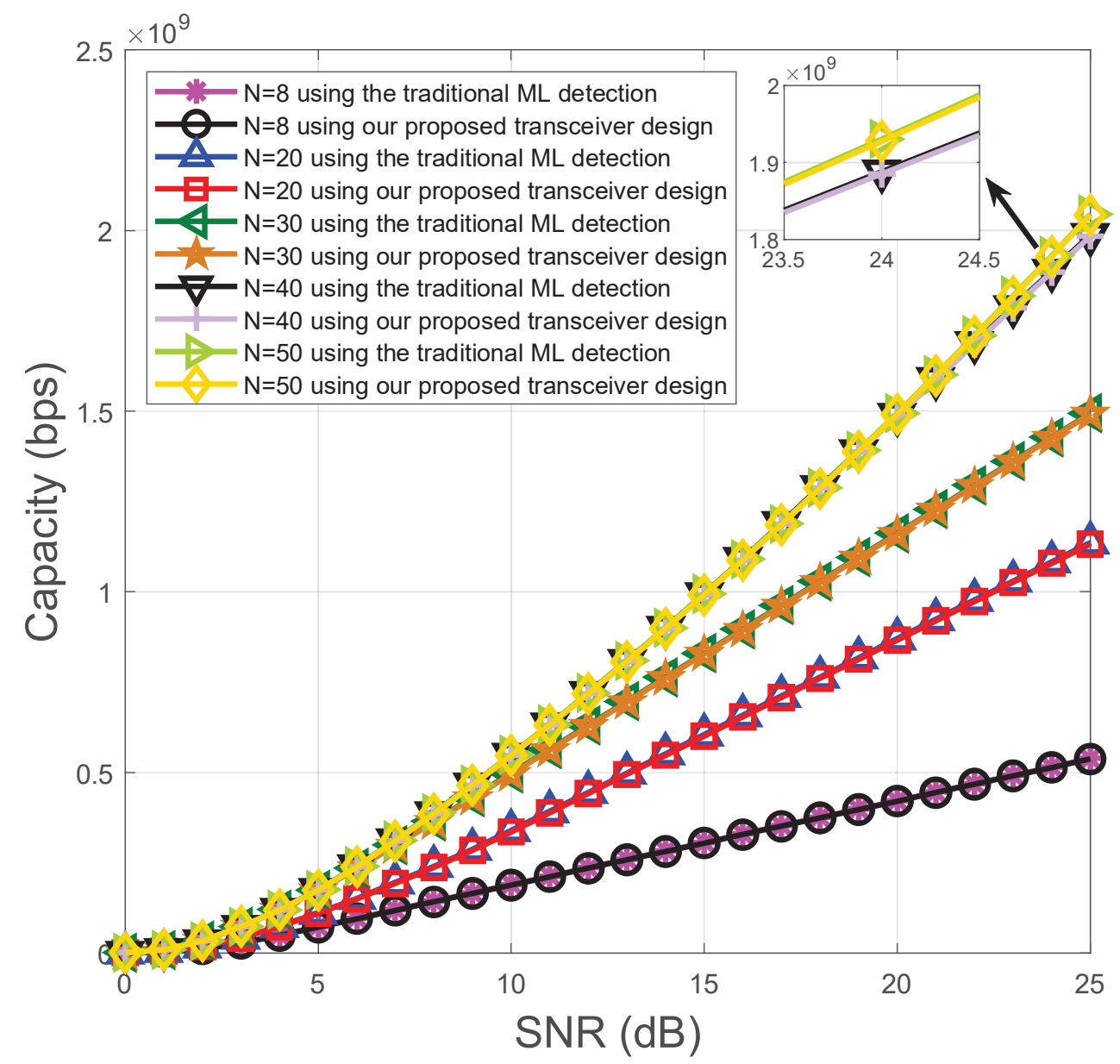

Fig. 9. Capacities for UCA based RIS-assisted MIMO system using our proposed transceiver design and using the traditional ML detection.

Fig. 8 illustrates the BERs versus rotation angles corresponding to the UCA based RIS-assisted MIMO system using our proposed transceiver design and using the traditional ML detection for the misaligned transceiver. Since the precoding matrix on the transmitter, the phase design matrices on the RIS UCAs, and the phase compensation matrix on the receiver are unitary, the BER for the UCA based RIS-assisted MIMO system using our proposed transceiver design is the same to that using the traditional ML detection. Generally, the BERs first decrease and then increase as the rotation angles increase.

Fig. 9 depicts the capacities of the UCA based RIS-assisted MIMO system using our proposed transceiver design and using the traditional ML detection for the misaligned transceiver scenario. Since the precoding matrix on the transmitter, the phase compensation matrices on the RIS UCAs, and the phase compensation matrix on the receiver are unitary, resulting in the channel properties keep unchanged, the capacities of the UCA based RIS-assisted MIMO system using our proposed transceiver design and using the traditional ML detection are the same. Due to multiplexing, the capacities increase as the number $N$ of antennas increases for the UCA based RIS-assisted MIMO system using our proposed transceiver design and using the traditional ML detection. However, the capacities for different antenna numbers $N$ do not change much when $N$ becomes large. This is because the channels become correlated when $N$ is large.

## VII. Conclusions

In this paper, we adopted two UCA based RISs between possibly misaligned transmitter and receiver to assist the transmission, where both transmitter and receiver are equipped with UCA antennas. We first investigated the channel model for such a system, where the transmitter, the RIS UCAs, and the receiver are not aligned. We developed an algorithm to obtain the ranges of rotation angles and reduce the impact of rotation angles on channel matrix. Also, we proposed a low complexity precoding scheme, phase designs for the RIS UCAs, and a phase compensation scheme such that the overall channel matrix from transmitter to receiver can be well approximated by a circulant matrix. Then, a fast symbol-wise ML detection scheme was proposed to recover the signals with low computational complexity. Simulation results were presented to validate the theoretical results.

## Appendix A
## Proof of Lemma 1

$$
\begin{aligned}
\frac{1}{d^{tr}_{mn}} &\approx \frac{1}{\sqrt{(d^{tr})^2+R_t^2+r_1^2+\frac{\mathcal{E}^{tr}_n+\mathcal{D}^{tr}_{mn}+\mathcal{F}^{tr}_m}{\sqrt{(d^{tr})^2+R_t^2+r_1^2}}}} \\
&= \frac{1}{\sqrt{(d^{tr})^2+R_t^2+r_1^2}} - \\
&\quad \frac{\frac{\mathcal{E}^{tr}_n+\mathcal{D}^{tr}_{mn}+\mathcal{F}^{tr}_m}{\sqrt{(d^{tr})^2+R_t^2+r_1^2}}}{\sqrt{(d^{tr})^2+R_t^2+r_1^2}\left(\sqrt{(d^{tr})^2+R_t^2+r_1^2}+\frac{\mathcal{E}^{tr}_n+\mathcal{D}^{tr}_{mn}+\mathcal{F}^{tr}_m}{\sqrt{(d^{tr})^2+R_t^2+r_1^2}}\right)} \\
&\overset{(a)}{=} \frac{1}{\sqrt{(d^{tr})^2+R_t^2+r_1^2}} - O\left[\left(\frac{1}{\sqrt{(d^{tr})^2+R_t^2+r_1^2}}\right)^2\right] \\
&\approx \frac{1}{\sqrt{(d^{tr})^2+R_t^2+r_1^2}},
\end{aligned} \tag{60}
$$

where Step $(a)$ is because $d^{tr} >> R_t$, $d^{tr} >> r_1$, and, from (8), we have

$$
\begin{aligned}
&\left|\frac{\mathcal{E}^{tr}_n+\mathcal{D}^{tr}_{mn}+\mathcal{F}^{tr}_m}{\sqrt{(d^{tr})^2+R_t^2+r_1^2}}\right| < \left|\frac{\mathcal{E}^{tr}_n+\mathcal{D}^{tr}_{mn}+\mathcal{F}^{tr}_m}{d^{tr}}\right| \\
&< \left|\frac{\mathcal{E}^{tr}_n}{d^{tr}}\right| + \left|\frac{\mathcal{D}^{tr}_{mn}}{d^{tr}}\right| + \left|\frac{\mathcal{F}^{tr}_m}{d^{tr}}\right| \\
&< \sqrt{5+2\sqrt{2}}R_t + \frac{R_t r_1}{d^{tr}} + r_1 < 3R_t + r_1.
\end{aligned} \tag{61}
$$

## Appendix B
## Proof of Lemma 2

To prove (42), the left hand side of (42) can be rewritten as:

$$
\begin{aligned}
&\sum_{m=1}^{N}\sum_{u=1}^{N} e^{-j\frac{2\pi}{\lambda}\left(\frac{\hat{\mathcal{D}}^{rr}_{ku}}{\sqrt{(d^{rr})^2+R_r^2+r_2^2}}+\frac{\hat{\mathcal{D}}^{Rr}_{um}}{\sqrt{(d^{Rr})^2+r_1^2+r_2^2}}+\frac{\hat{\mathcal{D}}^{tr}_{mn}}{\sqrt{(d^{tr})^2+R_t^2+r_1^2}}\right)} \\
&= \sum_{m=2}^{N}\sum_{u=2}^{N} e^{-j\frac{2\pi}{\lambda}\left(\frac{\hat{\mathcal{D}}^{rr}_{k(u-1)}}{\sqrt{(d^{rr})^2+R_r^2+r_2^2}}+\frac{\hat{\mathcal{D}}^{Rr}_{(u-1)(m-1)}}{\sqrt{(d^{Rr})^2+r_1^2+r_2^2}}+\frac{\hat{\mathcal{D}}^{tr}_{(m-1)n}}{\sqrt{(d^{tr})^2+R_t^2+r_1^2}}\right)} \\
&+ e^{-j\frac{2\pi}{\lambda}\left(\frac{\hat{\mathcal{D}}^{rr}_{kN}}{\sqrt{(d^{rr})^2+R_r^2+r_2^2}}+\frac{\hat{\mathcal{D}}^{Rr}_{NN}}{\sqrt{(d^{Rr})^2+r_1^2+r_2^2}}+\frac{\hat{\mathcal{D}}^{tr}_{Nn}}{\sqrt{(d^{tr})^2+R_t^2+r_1^2}}\right)} \\
&\sum_{u=2}^{N} e^{-j\frac{2\pi}{\lambda}\left(\frac{\hat{\mathcal{D}}^{rr}_{k(u-1)}}{\sqrt{(d^{rr})^2+R_r^2+r_2^2}}+\frac{\hat{\mathcal{D}}^{Rr}_{(u-1)N}}{\sqrt{(d^{Rr})^2+r_1^2+r_2^2}}+\frac{\hat{\mathcal{D}}^{tr}_{Nn}}{\sqrt{(d^{tr})^2+R_t^2+r_1^2}}\right)} \\
&+ \sum_{m=2}^{N} e^{-j\frac{2\pi}{\lambda}\left(\frac{\hat{\mathcal{D}}^{rr}_{kN}}{\sqrt{(d^{rr})^2+R_r^2+r_2^2}}+\frac{\hat{\mathcal{D}}^{Rr}_{N(m-1)}}{\sqrt{(d^{Rr})^2+r_1^2+r_2^2}}+\frac{\hat{\mathcal{D}}^{tr}_{(m-1)n}}{\sqrt{(d^{tr})^2+R_t^2+r_1^2}}\right)},
\end{aligned} \tag{62}
$$

In $\hat{\mathcal{D}}^{rr}_{ku}$, $\hat{\mathcal{D}}^{Rr}_{um}$, and $\hat{\mathcal{D}}^{tr}_{mn}$ of the left hand side of (42), angle $\psi_i = 2\pi(i-1)/N$ when $i$ is $k$, $u$, or $n$, and angle $\varphi_i = 2\pi(i-1)/N$ when $i$ is $u$ or $m$. In the right hand side of (62), terms $\hat{\mathcal{D}}^{rr}_{k(u-1)}$, $\hat{\mathcal{D}}^{Rr}_{(u-1)(m-1)}$, and $\hat{\mathcal{D}}^{tr}_{(m-1)n}$ can be derived in (63), (64), and (65), respectively, where $A^t$ and $A^r$ are defined in (41).

$$
\begin{aligned}
\hat{\mathcal{D}}^{rr}_{k(u-1)} &= -\frac{R_r r_2}{2}(1+\cos\vartheta^{rr}_y\cos\vartheta^{rr}_x)\cos(\psi_k-\varphi_{u-1}+\alpha_{R_r}-\alpha_{r2}-\chi^{rr}_1) \\
&= -\frac{R_r r_2}{2}(1+\cos\vartheta^{rr}_y\cos\vartheta^{rr}_x)\cos\left(\psi_k-\varphi_u+\alpha_{R_r}-\alpha_{r2}-\chi^{rr}_1+\frac{2\pi}{N}\right) \\
&= \frac{R_r r_2}{-2}(1+\cos\vartheta^{rr}_y\cos\vartheta^{rr}_x)\Big[\cos\frac{2\pi}{N}\cos(\psi_k-\varphi_u+\alpha_{R_r}-\alpha_{r2}-\chi^{rr}_1)-\sin\frac{2\pi}{N}\sin(\psi_k-\varphi_u+\alpha_{R_r}-\alpha_{r2}-\chi^{rr}_1)\Big] \\
&= -\frac{R_r r_2}{2}(1+\cos\vartheta^{rr}_y\cos\vartheta^{rr}_x)\Big[\cos(\psi_k-\varphi_u+\alpha_{R_r}-\alpha_{r2}-\chi^{rr}_1) \\
&\qquad -(1-\cos\frac{2\pi}{N})\cos(\psi_k-\varphi_u+\alpha_{R_r}-\alpha_{r2}-\chi^{rr}_1)-\sin\frac{2\pi}{N}\sin(\psi_k-\varphi_u+\alpha_{R_r}-\alpha_{r2}-\chi^{rr}_1)\Big] \\
&= \frac{R_r r_2(1+\cos\vartheta^{rr}_y\cos\vartheta^{rr}_x)}{-2}\Bigg[\cos(\psi_k-\varphi_u+\alpha_{R_r}-\alpha_{r2}-\chi^{rr}_1) \\
&\qquad -\sqrt{2-2\cos\frac{2\pi}{N}}\cos\left(\psi_k-\varphi_u+\alpha_{R_r}-\alpha_{r2}-\chi^{rr}_1-\arctan\frac{\sin(2\pi/N)}{1-\cos(2\pi/N)}\right)\Bigg] \\
&= \hat{\mathcal{D}}^{rr}_{ku}+A^r\sqrt{(d^{rr})^2+R_r^2+r_2^2}\cos\left(\psi_k-\varphi_u+\alpha_{R_r}-\alpha_{r2}-\chi^{rr}_1-\arctan\frac{\sin(2\pi/N)}{1-\cos(2\pi/N)}\right)
\end{aligned} \tag{63}
$$

$$\hat{\mathcal{D}}^{Rr}_{(u-1)(m-1)} = -\frac{r_1 r_2}{2}(1+\cos\vartheta^{Rr}_y\cos\vartheta^{Rr}_x)\cos\left(\psi_{u-1}-\varphi_{m-1}+\alpha_{r2}-\alpha_{r1}-\chi^{Rr}_1\right)$$
$$= -\frac{r_1 r_2}{2}(1+\cos\vartheta^{Rr}_y\cos\vartheta^{Rr}_x)\cos\left(\psi_{u}-\varphi_{m}+\alpha_{r2}-\alpha_{r1}-\chi^{Rr}_1\right) = \hat{\mathcal{D}}^{Rr}_{um} \tag{64}$$

$$\hat{\mathcal{D}}^{tr}_{(m-1)n} = -\frac{R_t r_1}{2}(1+\cos\vartheta^{tr}_y\cos\vartheta^{tr}_x)\cos\left(\psi_n-\varphi_{m-1}+\alpha_{R_t}-\alpha_{r1}-\chi^{tr}_1\right)$$
$$= -\frac{R_t r_1}{2}(1+\cos\vartheta^{tr}_y\cos\vartheta^{tr}_x)\cos\left(\psi_n-\varphi_{m}+\alpha_{R_t}-\alpha_{r1}-\chi^{tr}_1+\frac{2\pi}{N}\right)$$
$$= \frac{R_t r_1(1+\cos\vartheta^{tr}_y\cos\vartheta^{tr}_x)}{-2}\Bigg[\cos\left(\psi_n-\varphi_{m}+\alpha_{R_t}-\alpha_{r1}-\chi^{tr}_1\right)$$
$$-\sqrt{2-2\cos\frac{2\pi}{N}}\cos\left(\psi_n-\varphi_{m}+\alpha_{R_t}-\alpha_{r1}-\chi^{tr}_1-\arctan\frac{\sin(2\pi/N)}{1-\cos(2\pi/N)}\right)\Bigg]$$
$$= \hat{\mathcal{D}}^{tr}_{mn}+A^t\sqrt{(d^{tr})^2+R_t^2+r_1^2}\cos\left(\psi_n-\varphi_{m}+\alpha_{R_t}-\alpha_{r1}-\chi^{tr}_1-\arctan\frac{\sin(2\pi/N)}{1-\cos(2\pi/N)}\right) \tag{65}$$

Since $\psi_N=\psi_1+2\pi-2\pi/N$ and $\varphi_N=\varphi_1+2\pi-2\pi/N$, we have

$$\hat{\mathcal{D}}^{rr}_{kN}$$
$$=\frac{R_r r_2}{-2}(1+\cos\vartheta^{rr}_y\cos\vartheta^{rr}_x)\cos\left(\psi_k-\varphi_N+\alpha_{R_r}-\alpha_{r2}-\chi^{rr}_1\right)$$
$$=\frac{R_r r_2}{-2}(1+\cos\vartheta^{rr}_y\cos\vartheta^{rr}_x)\cos\left(\psi_k-\varphi_1+\alpha_{R_r}-\alpha_{r2}-\chi^{rr}_1+\frac{2\pi}{N}\right)$$
$$=\hat{\mathcal{D}}^{rr}_{k1}+A^r\sqrt{(d^{rr})^2+R_r^2+r_2^2}$$
$$\times\cos\left(\psi_k-\varphi_u+\alpha_{R_r}-\alpha_{r2}-\chi^{rr}_1-\arctan\frac{\sin(2\pi/N)}{1-\cos(2\pi/N)}\right), \tag{66}$$

$$\hat{\mathcal{D}}^{tr}_{Nn}$$
$$=\frac{R_t r_1}{-2}(1+\cos\vartheta^{tr}_y\cos\vartheta^{tr}_x)\cos\left(\psi_n-\varphi_N+\alpha_{R_t}-\alpha_{r1}-\chi^{tr}_1\right)$$
$$=\frac{R_t r_1}{-2}(1+\cos\vartheta^{tr}_y\cos\vartheta^{tr}_x)\cos\left(\psi_n-\varphi_1+\alpha_{R_t}-\alpha_{r1}-\chi^{tr}_1+\frac{2\pi}{N}\right)$$
$$=\hat{\mathcal{D}}^{tr}_{1n}+A^t\sqrt{(d^{tr})^2+R_t^2+r_1^2}$$
$$\times\cos\left(\psi_n-\varphi_m+\alpha_{R_t}-\alpha_{r1}-\chi^{tr}_1-\arctan\frac{\sin(2\pi/N)}{1-\cos(2\pi/N)}\right), \tag{67}$$

$$\hat{\mathcal{D}}^{Rr}_{(u-1)N}$$
$$=\frac{r_1 r_2}{-2}(1+\cos\vartheta^{Rr}_y\cos\vartheta^{Rr}_x)\cos\left(\psi_{u-1}-\varphi_N+\alpha_{r2}-\alpha_{r1}-\chi^{Rr}_1\right)$$
$$=\frac{r_1 r_2}{-2}(1+\cos\vartheta^{Rr}_y\cos\vartheta^{Rr}_x)\cos\left(\psi_{u}-\varphi_1+\alpha_{r2}-\alpha_{r1}-\chi^{Rr}_1\right)$$
$$=\hat{\mathcal{D}}^{Rr}_{u1}, \tag{68}$$

$$\hat{\mathcal{D}}^{Rr}_{N(m-1)}$$
$$=\frac{r_1 r_2}{-2}(1+\cos\vartheta^{Rr}_y\cos\vartheta^{Rr}_x)\cos\left(\psi_N-\varphi_{m-1}+\alpha_{r2}-\alpha_{r1}-\chi^{Rr}_1\right)$$
$$=\frac{r_1 r_2}{-2}(1+\cos\vartheta^{Rr}_y\cos\vartheta^{Rr}_x)\cos\left(\psi_1-\varphi_m+\alpha_{r2}-\alpha_{r1}-\chi^{Rr}_1\right)$$
$$=\hat{\mathcal{D}}^{Rr}_{1m}, \tag{69}$$

and

$$\hat{\mathcal{D}}^{Rr}_{NN}$$
$$=\frac{r_1 r_2}{-2}(1+\cos\vartheta^{Rr}_y\cos\vartheta^{Rr}_x)\cos\left(\psi_N-\varphi_N+\alpha_{r2}-\alpha_{r1}-\chi^{Rr}_1\right)$$
$$=\frac{r_1 r_2}{-2}(1+\cos\vartheta^{Rr}_y\cos\vartheta^{Rr}_x)\cos\left(\psi_1-\varphi_1+\alpha_{r2}-\alpha_{r1}-\chi^{Rr}_1\right)$$
$$=\hat{\mathcal{D}}^{Rr}_{11}. \tag{70}$$

Then, the left hand side of (62) can be rewritten as follows:

$$\sum_{m=1}^{N}\sum_{u=1}^{N}e^{-j\frac{2\pi}{\lambda}\left(\frac{\hat{\mathcal{D}}^{rr}_{ku}}{\sqrt{(d^{rr})^2+R_r^2+r_2^2}}+\frac{\hat{\mathcal{D}}^{Rr}_{um}}{\sqrt{(d^{Rr})^2+r_1^2+r_2^2}}+\frac{\hat{\mathcal{D}}^{tr}_{mn}}{\sqrt{(d^{tr})^2+R_t^2+r_1^2}}\right)}$$
$$=\sum_{m=2}^{N}\sum_{u=2}^{N}\Bigg[e^{-j\frac{2\pi}{\lambda}\left(\frac{\hat{\mathcal{D}}^{rr}_{ku}}{\sqrt{(d^{rr})^2+R_r^2+r_2^2}}+\frac{\hat{\mathcal{D}}^{Rr}_{um}}{\sqrt{(d^{Rr})^2+r_1^2+r_2^2}}+\frac{\hat{\mathcal{D}}^{tr}_{mn}}{\sqrt{(d^{tr})^2+R_t^2+r_1^2}}\right)}$$
$$\times e^{-j\frac{2\pi}{\lambda}A^t\cos\left(\psi_n-\varphi_m+\alpha_{R_t}-\alpha_{r1}-\chi^{tr}_1-\arctan\frac{\sin\frac{2\pi}{N}}{1-\cos\frac{2\pi}{N}}\right)}$$
$$\times e^{-j\frac{2\pi}{\lambda}A^r\cos\left(\psi_k-\varphi_u+\alpha_{R_r}-\alpha_{r2}-\chi^{rr}_1-\arctan\frac{\sin\frac{2\pi}{N}}{1-\cos\frac{2\pi}{N}}\right)}\Bigg]$$
$$+\sum_{u=2}^{N}\Bigg[e^{-j\frac{2\pi}{\lambda}\left(\frac{\hat{\mathcal{D}}^{rr}_{ku}}{\sqrt{(d^{rr})^2+R_r^2+r_2^2}}+\frac{\hat{\mathcal{D}}^{Rr}_{u1}}{\sqrt{(d^{Rr})^2+r_1^2+r_2^2}}+\frac{\hat{\mathcal{D}}^{tr}_{1n}}{\sqrt{(d^{tr})^2+R_t^2+r_1^2}}\right)}$$
$$\times e^{-j\frac{2\pi}{\lambda}A^t\cos\left(\psi_n-\varphi_1+\alpha_{R_t}-\alpha_{r1}-\chi^{tr}_1-\arctan\frac{\sin\frac{2\pi}{N}}{1-\cos\frac{2\pi}{N}}\right)}$$
$$\times e^{-j\frac{2\pi}{\lambda}A^r\cos\left(\psi_k-\varphi_u+\alpha_{R_r}-\alpha_{r2}-\chi^{rr}_1-\arctan\frac{\sin\frac{2\pi}{N}}{1-\cos\frac{2\pi}{N}}\right)}\Bigg]$$
$$+\sum_{m=2}^{N}\Bigg[e^{-j\frac{2\pi}{\lambda}\left(\frac{\hat{\mathcal{D}}^{rr}_{k1}}{\sqrt{(d^{rr})^2+R_r^2+r_2^2}}+\frac{\hat{\mathcal{D}}^{Rr}_{1m}}{\sqrt{(d^{Rr})^2+r_1^2+r_2^2}}+\frac{\hat{\mathcal{D}}^{tr}_{mn}}{\sqrt{(d^{tr})^2+R_t^2+r_1^2}}\right)}$$
$$\times e^{-j\frac{2\pi}{\lambda}A^t\cos\left(\psi_n-\varphi_m+\alpha_{R_t}-\alpha_{r1}-\chi^{tr}_1-\arctan\frac{\sin\frac{2\pi}{N}}{1-\cos\frac{2\pi}{N}}\right)}$$
$$\times e^{-j\frac{2\pi}{\lambda}A^r\cos\left(\psi_k-\varphi_1+\alpha_{R_r}-\alpha_{r2}-\chi^{rr}_1-\arctan\frac{\sin\frac{2\pi}{N}}{1-\cos\frac{2\pi}{N}}\right)}\Bigg]$$
$$+\Bigg[e^{-j\frac{2\pi}{\lambda}\left(\frac{\hat{\mathcal{D}}^{rr}_{k1}}{\sqrt{(d^{rr})^2+R_r^2+r_2^2}}+\frac{\hat{\mathcal{D}}^{Rr}_{11}}{\sqrt{(d^{Rr})^2+r_1^2+r_2^2}}+\frac{\hat{\mathcal{D}}^{tr}_{1n}}{\sqrt{(d^{tr})^2+R_t^2+r_1^2}}\right)}$$

$$
\begin{aligned}
&\times e^{-j\frac{2\pi}{\lambda}A^t\cos\left(\psi_n-\varphi_1+\alpha_{R_t}-\alpha_{r1}-\chi_1^{tr}-\arctan\frac{\sin\frac{2\pi}{N}}{1-\cos\frac{2\pi}{N}}\right)}\\
&\times e^{-j\frac{2\pi}{\lambda}A^r\cos\left(\psi_k-\varphi_1+\alpha_{R_r}-\alpha_{r2}-\chi_1^{rr}-\arctan\frac{\sin\frac{2\pi}{N}}{1-\cos\frac{2\pi}{N}}\right)}\Bigg]\\
&=\sum_{m=1}^{N}\sum_{u=1}^{N}\Bigg[e^{-j\frac{2\pi}{\lambda}\left(\frac{\hat{\mathcal{D}}_{ku}^{rr}}{\sqrt{(d^{rr})^2+R_r^2+r_2^2}}+\frac{\hat{\mathcal{D}}_{um}^{Rr}}{\sqrt{(d^{Rr})^2+r_1^2+r_2^2}}+\frac{\hat{\mathcal{D}}_{mn}^{tr}}{\sqrt{(d^{tr})^2+R_t^2+r_1^2}}\right)}\\
&\times e^{-j\frac{2\pi}{\lambda}A^t\cos\left(\psi_n-\varphi_m+\alpha_{R_t}-\alpha_{r1}-\chi_1^{tr}-\arctan\frac{\sin\frac{2\pi}{N}}{1-\cos\frac{2\pi}{N}}\right)}\\
&\times e^{-j\frac{2\pi}{\lambda}A^r\cos\left(\psi_k-\varphi_u+\alpha_{R_r}-\alpha_{r2}-\chi_1^{rr}-\arctan\frac{\sin\frac{2\pi}{N}}{1-\cos\frac{2\pi}{N}}\right)}\Bigg].
\end{aligned}
\tag{71}
$$

Therefore, from (71), we obtain (42).

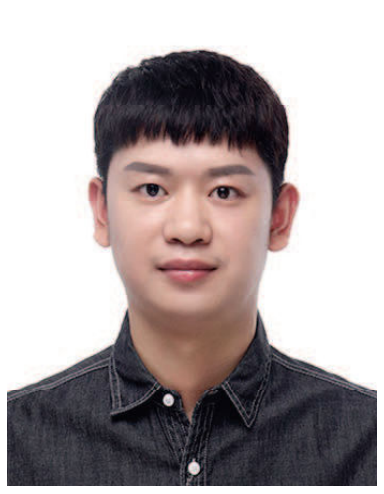

**Haiyue Jing** received the B.S. degree in telecommunication engineering from Xidian University, China, in 2017. He is currently pursuing the Ph.D. degree in telecommunication engineering at Xidian University. His research interests focus on B5G/6G wireless networks, OAM based wireless communications, and LOS MIMO wireless communications.

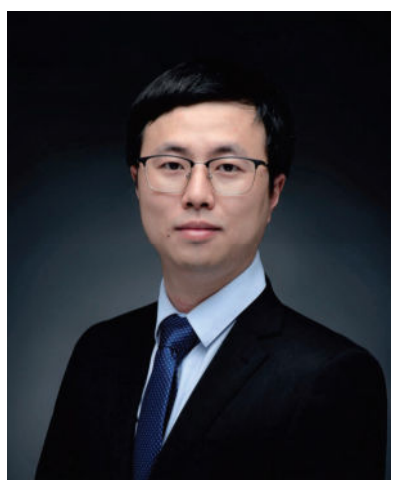

**Wenchi Cheng** (Senior Member, IEEE) received the B.S. and Ph.D. degrees in telecommunication engineering from Xidian University, Xian, China, in 2008 and 2013, respectively. He was a Visiting Scholar with the Department of Electrical and Computer Engineering, Texas A&M University, College Station, TX, USA, from 2010 to 2011. He is currently a Full Professor with Xidian University. His current research interests include B5G/6G wireless networks, emergency wireless communications, and orbital-angular-momentum-based wireless communications.

He has published more than 100 international journal and conference papers in IEEE JOURNAL ON SELECTED AREAS IN COMMUNICATIONS, IEEE magazines, IEEE TRANSACTIONS, IEEE INFOCOM, GLOBECOM, and ICC. He received the IEEE ComSoc Asia-Pacific Outstanding Young Researcher Award in 2021, the URSI Young Scientist Award in 2019, the Young Elite Scientist Award of CAST, and four IEEE journal/conference best papers. He has served or serving as the Wireless Communications Symposium Co-Chair for IEEE ICC 2022 and IEEE GLOBECOM 2020, the Publicity Chair for IEEE ICC 2019, the Next Generation Networks Symposium Chair for IEEE ICCC 2019, and the Workshop Chair for IEEE ICC 2019/IEEE GLOBECOM 2019/INFOCOM 2020 Workshop on Intelligent Wireless Emergency Communications Networks. He has served or serving as an Associate Editor for IEEE SYSTEMS JOURNAL, IEEE COMMUNICATIONS LETTERS, and IEEE WIRELESS COMMUNICATIONS LETTER.

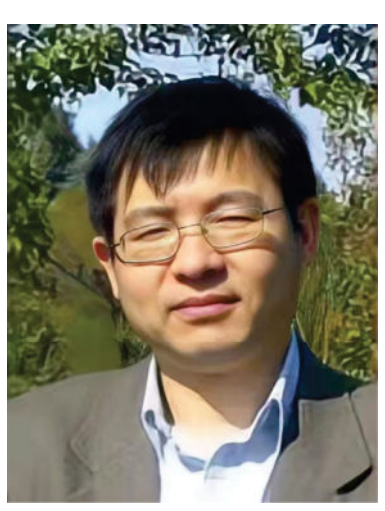

**Xiang-Gen Xia** (Fellow, IEEE) received the B.S. degree in mathematics from Nanjing Normal University, Nanjing, China, in 1983, the M.S. degree in mathematics from Nankai University, Tianjin, China, in 1968, and the Ph.D. degree in electrical engineering from the University of Southern California, Los Angeles, CA, USA, in 1992.

He was a Senior/Research Staff Member with the Hughes Research Laboratories, Malibu, CA, from 1995 to 1996. In 1996, he joined the Department of Electrical and Computer Engineering, University of Delaware, Newark, DE, USA, where he is currently the Charles Black Evans Professor. He has authored the book Modulated Coding for Intersymbol Interference Channels (Marcel Dekker, New York, 2000). His current research interests include space-time coding, MIMO and OFDM systems, digital signal processing, and synthetic aperture radar (SAR) and ISAR imaging.

Dr. Xia received the National Science Foundation (NSF) Faculty Early Career Development (CAREER) Program Award in 1997, the Office of Naval Research (ONR) Young Investigator Award in 1998, and the Outstanding Overseas Young Investigator Award from the National Nature Science Foundation of China in 2001. He received the 2019 Information Theory Outstanding Overseas Chinese Scientist Award from the Information Theory Society of the Chinese Institute of Electronics. He is the Technical Program Chair of the Signal Processing Symposium, Globecom 2007, in Washington, DC, and the General Co-Chair of the ICASSP 2005 in Philadelphia. He has served as an Associate Editor for numerous international journals, including IEEE Transactions on Signal Processing, IEEE Transactions on Wireless Communications, IEEE Transactions on Mobile Computing, and IEEE Transactions on Vehicular Technology.